\documentclass[
 reprint,
 amsmath,amssymb,
 aps,
]{revtex4-2}
\usepackage{amsmath,amssymb,amsfonts}
\usepackage{placeins}
\usepackage{hyperref}
\usepackage{braket}
\usepackage{orcidlink}
\usepackage{dsfont}
\usepackage{xcolor}
\usepackage{graphicx} 

\date{April 2025}

\begin{document}

\title{Polaronic Effect in High-Harmonic Generation}

\author{Gabriel C\'aceres-Aravena\,\orcidlink{0000-0003-1411-7026}}
\email{gabriel.caceres-aravena@uni-rostock.de}
\affiliation{
 Institute of Physics, University of Rostock, 18051 Rostock, Germany
}
\author{Dieter Bauer\,\orcidlink{0000-0003-4680-5560}}
 \email{dieter.bauer@uni-rostock.de}
\affiliation{
Institute of Physics, University of Rostock, 18051 Rostock, Germany
}

\begin{abstract}
We investigate High-Harmonic Generation (HHG) in the Su-Schrieffer-Heeger (SSH) chain with electron-phonon coupling modeled via the Holstein interaction. The system dynamics are simulated using the tight-binding approximation, with local phonons approximated as quantum harmonic oscillators. Phononic degrees of freedom significantly expand the Hilbert space dimension
. This interaction modifies the eigenenergy spectrum by introducing new states within previously existing gaps, enhancing the harmonic yield through additional allowed transitions.
\end{abstract}

                              
\maketitle

\section{Introduction}

High-Harmonic Generation (HHG) spectroscopy is a powerful technique in condensed matter physics for probing ultrafast electron dynamics driven by intense, ultrashort laser pulses.  
In HHG, a strong laser field accelerates electrons in a material, leading to the emission of radiation containing harmonics of the driving laser frequency \cite{goulielmakis_high_2022,PhysRevA.49.2117}. 
Initially studied in gases--enabling attosecond pulse generation \cite{PhysRevLett.71.1994}--HHG opened new possibilities for investigating ultrafast dynamics \cite{attosecond}. 
The first experimental observations of HHG in bulk crystals \cite{Ghimire2010} provided insights into the electronic structure of solids. 
Subsequent studies revealed that the harmonics generated in solids encode information about the material \cite{WOS:000454733100012}, including its band structure \cite{structures}, topological phase \cite{berry}, and other properties. 
This occurs because the accelerated electrons move through the material according to its electronic bands under the influence of the laser field. Therefore, any modifications to the electronic band structure can affect the electron dynamics and, consequently, alter the HHG spectrum. 
As a result, HHG has emerged as a promising technique for studying the band structure and dynamical properties of materials  \cite{smirnova_high_2009,BauerHansen2018,JuerssBauer2019,Pooyan_Bauer_2025,Ghimire2010,Schubert2014,structures,Hohenleutner2015,Luu2015,Ndabashimiye2016,Langer2017,PhysRevLett.118.087403,You2017,Zhang2018,Vampa2018,Baudisch2018,Garg2018}. 
In the recent years, there has been a growing interest in the study of topological condensed matter and strong-field HHG theoretically \cite{BauerHansen2018,JuerssBauer2019,PhysRevA.99.053402,PhysRevA.102.053112,PhysRevB.96.075409,PhysRevX.13.031011}  and experimentally \cite{luu_measurement_2018,reimann_subcycle_2018,korobenko_high-harmonic_2021,https://doi.org/10.1002/adom.202203070,10.1063/5.0191184,WOS:000969760700001}. 
A key focus is to understand how the HHG spectrum is influenced by additional interactions, such as electron-electron (e-e), electron-phonon (e-ph) interactions, many-body effects, or topological phases. 

Significant progress has been made in the study of the many-body HHG, with most discussions centered in the e-e interaction \cite{PhysRevB.69.245311} and the topological characteristics of the harmonic spectra \cite{BauerHansen2018}. Nevertheless, little attention has been given to the e-ph interactions, despite the fact that several classes of materials that exhibit e-ph interaction such as in crystals \cite{Austin01112001,https://doi.org/10.1002/pssb.19680270144} or other materials \cite{franchini_polarons_2021,CREVECOEUR1970783,Stoneham_2007,coropceanu_charge_2007,WOS:000352259800014,de_sio_tracking_2016,PhysRevLett.88.247202,teresa_evidence_1997,PhysRevLett.89.097205,PhysRevLett.77.904,zhao_evidence_1997,C7TC00366H,doi:10.1126/sciadv.1701217,10.1063/1.5025907,WOS:000439573400010,PhysRevLett.108.116403,10.1063/5.0076070,morita_models_2023}. 
The e-ph interaction is a fundamental aspect of condensed matter physics, playing a key role in determining materials properties and influencing a wide variety of phenomena, from superconductivity to phase transitions. In this context, the Holstein model \cite{HOLSTEIN} is a prototypical framework for describing e-ph interactions \cite{PhysRevLett.89.196401,PhysRevB.56.4513,PhysRevB.48.6302,PhysRevB.60.14092}, where electrons are coupled to Einstein phonons (i.e., dispersionless local phonons). 
A key feature of the Holstein model is that the local e-ph interaction leads to the formation of a bound e-ph state. This results in the creation of a quasiparticle known as a polaron \cite{Holstein_Brink}. 

In the literature, the Holstein model has been widely used, often in conjunction with other models, to describe various types of many-body interactions in condensed matter systems. This combined models provide a quantum mechanical framework for studying rich phenomena such as superconductivity, antiferromagnetism, and charge-ordered phases \cite{PhysRevB.88.125126}, as well as for simulating these effects in quantum devices \cite{PhysRevB.107.165155}. 
In the context of HHG, the Holstein model has been employed to investigate dynamical field theory and scattering effects \cite{WOS:000314516500003}, as well as phonon-mediated interactions in superconductors \cite{WOS:000391009300002} and Mott insulators \cite{doi:10.7566/JPSJ.84.054403}. However, despite these important advances, the specific role of e-ph coupling in HHG spectra remains relatively underexplored, particularly regarding how phonon dynamics may influence ultrafast nonlinear optical responses.

In this work we study the effect of the Holstein polaron \cite{HOLSTEIN} on HHG. To this end, we perform numerical calculations within the tight-binding approximation to simulate the coupled electronic and phononic dynamics under a strong laser field. 
We find that including phonon degrees of freedom causes the Hilbert space to grow exponentially. As a result, we are constrained to simulate only relatively short systems with one electron and a few phonon energy levels. 
Therefore, we use the Su–Schrieffer–Heeger (SSH) chain \cite{PhysRevLett.42.1698}, a minimal lattice that exhibits a band gap and requires only a few sites per unit cell. 
The SSH chain has been previously studied in the context of HHG using time-dependent density-functional theory \cite{BauerHansen2018}, as well as within tight-binding models without e-e interaction \cite{JuerssBauer2019} and with the Hubbard e-e interaction \cite{Pooyan_Bauer_2025}. 
However, the effect of different e-ph couplings of Holstein type on HHG in the SSH model has, to best of our knowledge, not been explored. 
Understanding how phonon-mediated interactions modify HHG spectra can shed light on the ultrafast response of correlated materials and help identify new mechanisms for spectral control.

The main body of this work is structured into four sections: (i) The introduction. (ii) The theoretical framework, beginning with the Hamiltonian of the SSH Holstein system and basis construction. We then introduce the methodology used for the HHG spectroscopy. (iii) Simulation results, including electron and phonon dynamics. We analyze the spectra across various values of the e-ph coupling to explore its effect on harmonic generation. (iv) We presents our conclusions with a summary of our main findings.

\section{Theory}

\subsection{The Holstein SSH model}

\begin{figure}[h!]\centering
	\includegraphics[width=0.47\textwidth]{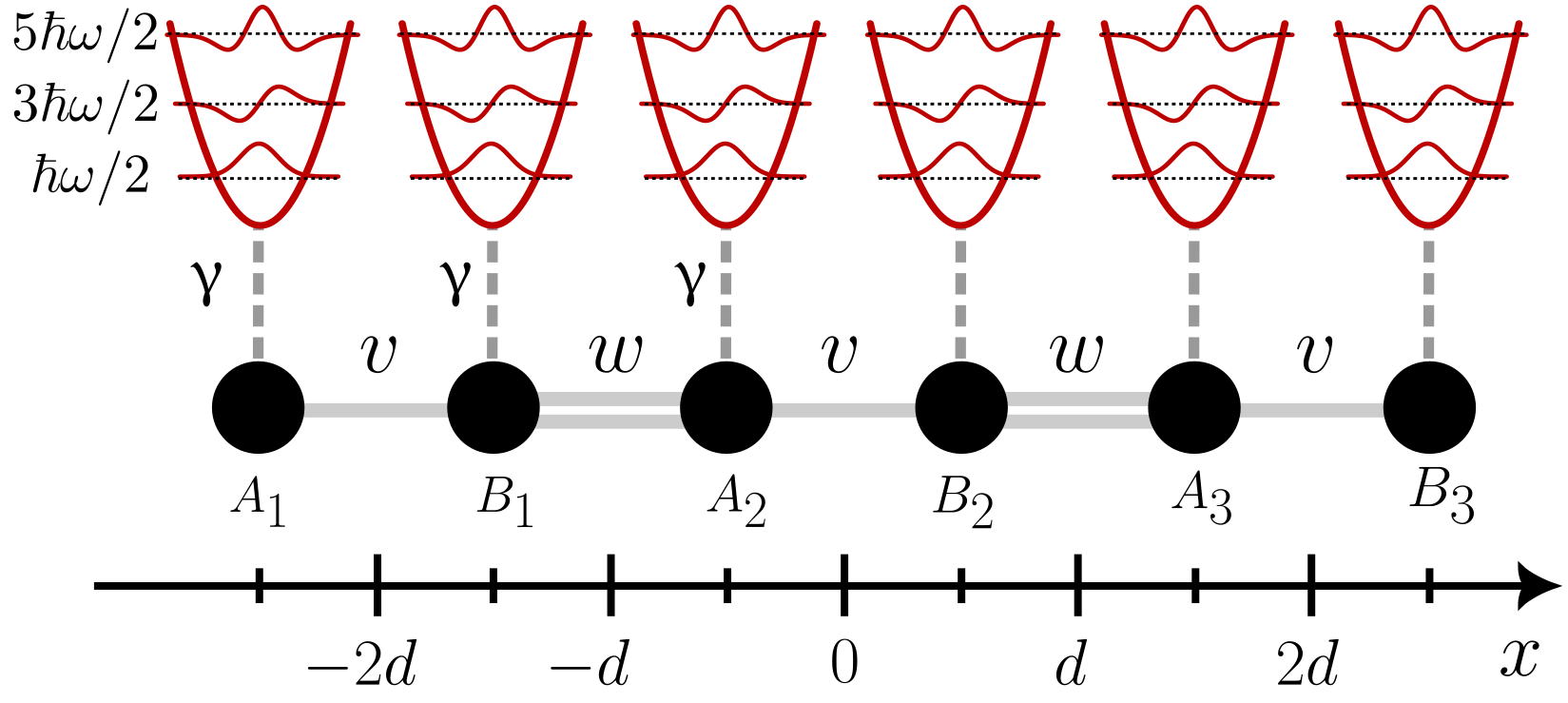}
	\caption[Simple model]{Sketch of the model. Sites are represented by black circles. The hoppings $v$ and $w$ are shown as single and double gray lines, respectively. The average distance between ions is $d$. The e-ph interaction $\gamma$ is indicated by dashed gray lines. Local phonons are illustrated in red.}\label{simple_model} 
\end{figure}

Consider the SSH lattice as sketched in the Fig. \ref{simple_model}, which consists of two sites per unit cell: $A$ and $B$. There are two types of electron-electron hoppings: $v<0$ represented by a single line, and $w<0$ represented by a double line. The average distance between ions is $d$, and a small displacement in the ions leads to dimerization, however, this displacement is approximately two orders of magnitude smaller than $d$. 
We define $D$ as the total length of the chain and set the reference frame such that the first site is located at $-D/2$ and the last site at $D/2$. 

The Holstein SSH Hamiltonian is given by:
\begin{align} \label{eq:holstein_hamiltonian}
	\hat H_\text{H} &= \hat H_e + \hat H_{ph} + \hat H_{e-ph} \ ,
\end{align}
where $\hat H_e$ is the electron kinetic term, $\hat H_{ph}$ represents the phonon energy, and $\hat H_{e-ph}$ denotes the e-ph interaction. 

The $\hat H_e$ term is given by:
\begin{align}\label{eq:electron_kinetic}
    \hat H_e &= \sum_{n=1}^N \left( v\;\hat c_{A,n}^\dagger \hat c_{B,n}+w\;\hat c_{A,n+1}^\dagger \hat c_{B,n} \right) +h.c.\ ,
\end{align}
where $\hat c_{i,n}$ and $\hat c_{i,n}^\dagger$ are the electron annihilation and creation operator at site  $i=\{A,B\}$ in the $n$th unit cell, and $N$ is the total number of unit cells. 

In the Holstein model, each site hosts localized quantum phonon states described by the solutions of the quantum harmonic oscillator \cite{HOLSTEIN,Holstein_Brink}. 
In the sketch shown in Fig.\ref{simple_model}, these localized phonon states are illustrated in red as quantized energy levels of the harmonic oscillator, while the e-ph hopping $\gamma<0$ is depicted with dashed gray lines. 
In this manner $\hat H_{ph}$ is:
\begin{align}\label{eq:phonon_energy}
    \hat H_{ph} = \hbar \omega_{ph} \sum_{n=1}^{N} \sum_{i=\{A,B\}} \left( \hat b_{i,n}^\dagger \hat b_{i,n} + \frac{1}{2} \right)\ ,
\end{align}
where $\hbar\omega_{ph}$ is the phonon energy, $\hat b_{i,n}$, $\hat b_{i,n}^\dagger$ are the annihilation and creation operators of local phonons ($\hbar=1$). 

The $\hat H_{e-ph}$ term that meadiates the e-ph interaction is 
\begin{align}\label{eq:electron_phonon_coupling}
    \hat H_{e-ph} = \sum_{n=1}^{N}\sum_{i=\{A,B\}} \gamma\; \hat n_{e,i,n} (\hat b_{i,n}^\dagger + \hat b_{i,n}) \ ,
\end{align}
where $\hat n_{e,i,n}$ is the electron number operator. 
The inclusion of the e-ph interaction produces a decrease in the ground state energy when a polaron forms, a hypothesis we will examine later in this work.

\subsection{Hilbert space}

The Hilbert space of the e-ph system is constructed as $\mathcal{H}_H=\mathcal{H}_e\otimes \mathcal{H}_{ph}$, where  $\mathcal H_e$ and $\mathcal H_{ph}$ denote the Hilbert space of the electronic and phononic subsystems, respectively. 
The phonon number per site is truncated to $L$ levels for numerical tractability, and we use one electron. For $N$ unit cells, $\dim(\mathcal H_{ph}) = L^{2N}$, and $\dim(\mathcal H_{e}) = 2N$, and then the total dimension $\mathcal{N}=\dim\mathcal H_H=\dim(\mathcal H_{e}\otimes \mathcal H_{ph})=2N(L)^{2N}$ 
To limit $\mathcal{N}$, we use $L=3$ and $N=3$ (as shown in Fig. \ref{simple_model}) and assume low-energy phonons dominate low-lying excitations, justifying a small $L$. We verify convergence of spectra with increasing $L$.
We write a general state as:
\begin{align}
    \ket{\psi} = \sum_{n_{A_1},...,n_{B_N},r}  \alpha_{(n_{A_1},...,n_{B_N},r)}\ket{n_{A_1},...,n_{B_N}}_{ph} \ket{r}_{e}  \ ,
\end{align}
where $r$ labels the site in which the electron is localized, and $n_f$ denotes the number of phonons at the $f$th site, $\alpha_{(n_{A_1},...,n_{B_N},r)}$ is the normalized probability amplitude of the basis element $\ket{n_{A_1},n_{B_1},...,n_{B_N}}_{ph} \ket{r}_{e}$.

\subsection{Coupling parameters}

All reported parameters are given in atomic units (a.u.). 
We perform simulations using the electron hopping parameters $v=-0.073$ and $w=-0.104$, which are similar values as in Ref. \cite{PhysRevLett.42.1698}. 
Phonons are more easily excited when $\omega_{ph}$ is small compared to the phononless transition energies and $|\gamma|$. In this case, a larger cutoff $L$ is needed to accurately describe the phononic states, leading to an exponential increase in the basis and matrix size, this is an issue we aim to avoid. 
Conversely, if $\omega_{ph}$ is large compared to the phononless transition energies and $|\gamma|$, then fewer phonons are excited, and the low-energy states closely resemble the phononless case. 
Therefore, we seek a regime for $\omega_{ph}$ and $\gamma$ that does not require a large $L$ while still allowing the low-energy states to differ noticeably from the phononless scenario. 
To achieve this, we fix $\omega_{ph}$ and vary $\gamma$, ensuring that $\omega_{ph}$ is neither too small or too large relative to the phononless transition energies and $\gamma$. 
Therefore, we select $\omega_{ph}=0.036$, which is of the same magnitude order as the electron parameters. 
For the e-ph coupling $\gamma$ we initially use the value 
$\gamma=-0.025$, but this value is varied to explore the parameter space. 
For the laser parameters, we set the number of cycles to $n_{cyc}=5$, the peak vector potential to $A_0=0.183$, and the laser frequency to $\omega_L=0.002$.

\subsection{Energy levels}

Only a few key eigenstates are populated by the laser, as harmonic peaks mainly arise from ground state transitions to select excited states within the harmonic energy range \cite{Pooyan_Bauer_2025}. Thus, we focus on harmonics up to the $40$th order. 
To improve computational efficiency, we restrict the eigenstate basis to those with eigenenergies within this region of interest. 
We determine the harmonic order corresponding to an eigenenergy $\varepsilon$ using the relation $(\varepsilon-\varepsilon_\mathrm{gs})/\omega_{L}$, where $\omega_L$ is the laser frequency. 
We set the laser frequency $\omega_L$ to around 20 times the energy gap between the ground state and the first excited state for $L=1$. 
For each $L$, we compute the energy spectrum, then we use the $\mathcal{N}_R$ lowest eigenstates covering up to at least the $40$th harmonic order.

\begin{figure}
    \centering
    \includegraphics[width=1.0\linewidth]{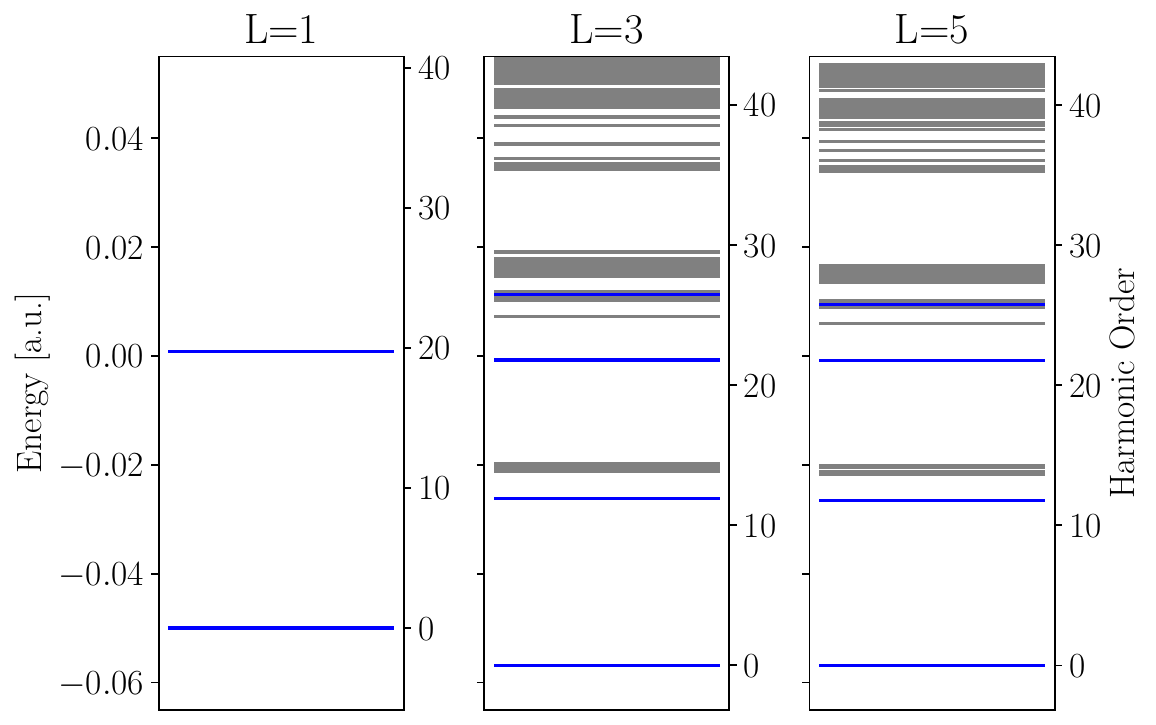}
    \caption{The energy levels for phonon cutoffs $L=\{1,3,5\}$. Blue indicates strong transitions from the ground state. Parameters used $v=-0.073$, $w=-0.104$ and phonon parameters $\omega_{ph}=0.036$, $\gamma=-0.025$ (all in a.u.).}
    \label{fig:energy-levels}
\end{figure}

Figure \ref{fig:energy-levels} shows the energy levels for $L=\{1,3,5\}$, up to the $40$th harmonic order. We use $\mathcal{N}_R=6$ for $L=1$, $\mathcal{N}_R=1500$ for $L=3$ and $\mathcal{N}_R=1700$ for $L=5$. We notice that for the case $L=5$ the new states appear at higher energies and with $\mathcal{N}_R=1700$ we cover the region of interest. 
Highlighted in blue are the ground state and excited states significantly populated via transitions from it (see Eq.(\ref{eq:transition})). 
For $L=1$, the local phonons can only occupy their ground state, so the phonon creation and annihilation operators cannot induce transitions, and e-ph interaction is suppressed. As a result, this case is equivalent to the phononless scenario. 
For $L=3$, the phonon operators allow e-ph interaction, and we observe in Fig.\ref{fig:energy-levels} that the ground state energy decreases compared to the $L=1$ case. This is a manifestation of the e-ph binding energy discussed earlier. 
For $L=5$, the ground state energy is similar to $L=3$, indicating that the ground state energy has begun to converge with respect to $L$. This observation supports the ansatz introduced earlier, which states that low-energy local phonons play a more significant role in the dynamics of low energy states. 
The supplementary material includes electron and phonon distributions of the most relevant states for $L=3$ highlighted in blue in Fig.\ref{fig:energy-levels}.

\subsection{High-Harmonic Generation}

To perform the high-harmonic simulation, we add a laser interaction term to the Hamiltonian. We consider a classical laser field polarized along the chain direction, treated within the dipole approximation in the length gauge. 
Then the Hamiltonian is:
\begin{align}\label{eq:fullH}
    \hat H(t) = \hat H_H +  E(t)\hspace{0.05cm} \hat x_e\ ,
\end{align}
where $\hat H_H$ is presented in Eq.(\ref{eq:holstein_hamiltonian}), 
$E(t)$ is the electric field amplitude and $\hat x_e$ is the position operator of electrons that we calculate as
\begin{align}
\hat x_e\approx \sum_{r=0}^{5}d( r-2.5)\hat n_{e,r}    \ ,
\end{align}
where $d=2$ (a.u.) is the average distance between ions. 

We simulate the system over the time interval $0\leq t\leq t_f=2\pi n_{cyc}/\omega_L$, using the electric field defined by:
\begin{align}
E=-\frac{\partial A(t) }{\partial t},\;
        A(t)=-A_0\sin^2\left(\frac{\omega_L t }{2 n_{cyc}}\right) \sin(\omega_L t ) , 
\end{align}
where $\omega_L$ is the laser frequency, $n_{cyc}$ is the number of field cycles and $A_0$ is the field amplitude.

Using the Hamiltonian in Eq.(\ref{eq:fullH}) we solve the time-dependent Schr\"odinger equation:
\begin{align}\label{eq1}
    i\partial_t \ket{\psi(t)}=\hat H(t) \ket{\psi(t)}\ .
\end{align}
We express the general state $\ket{\psi}$ as a linear combination of the eigenstates $\ket{\phi_n}$ of $H_H$ as follows
\begin{align}\label{eq:psi_as_eigenvectors}
    \ket{\psi(t)} = \sum_{n=0}^{\mathcal{N}-1 } a_n(t) \ket{\phi_n}\ .
\end{align}

The time dependence is contained in the electric field amplitude $E(t)$, while the braket forms a time invariant term that mediates the transition between states, which we define as the transition matrix $T$, and we also define the ground state transition vector $\vec T_\mathrm{gs}$ as: 
\begin{align}\label{eq:transition}
    T_{m n}= \bra{\phi_m} \hat x_e \ket{\phi_n} \ ,\hspace{0.2cm} [\vec T_\mathrm{gs}]_m = T_{\mathrm{gs},m}\ .
\end{align}
Where we have defined $\vec T_\mathrm{gs}$ due to the ground state transitions are more important because of the initial condition. 

The previous expansion in the time-dependent Schrödinger equation in Eq.(\ref{eq1}) leads to:
\begin{align}\label{eq:eq_sim}
i\partial_t a_m(t)= \varepsilon_m a_m(t) + \sum_{n=0}^{\mathcal{N}_R-1} a_n(t) E(t) T_{mn}  \ ,
\end{align}
where $\varepsilon_m$ is the eigenvalue of the eigenstate $\ket{\phi_m}$, 
and $0\leq m < \mathcal N_R$. 
The term inside the sum mediates the transition between the states $\phi_n$ and $\phi_m$. 
We solve these ordinary differential equations using the RK4 method, with the ground state $a_n(t=0)=\delta_{\mathrm{gs},n}$, as the initial condition, thus obtaining $a_n(t)$.

To compute the emitted harmonic yield, we employ a semi-classical approach where the radiated light is proportional to the electron acceleration. 
We calculate the electron position expectation value, apply a three point discrete derivative to calculate the acceleration, Then we use a Hanning window and perform a Fast Fourier Transform (FFT) to obtain the electron oscillation frequency. Finally, we take the absolute square and compute the base 10 logarithm as follows:
\begin{align}
   Y(\omega)= \log_{10} \left(\left|FFT\left(Hann\left( \frac{d^2 \langle \hat x_e \rangle}{dt^2} \right)\right)\right|^2\right)\ ,
\end{align}
where $\omega$ is the frequency obtained from the FFT. 
We normalize the yield to the intensity of the fundamental harmonic using $Y_N(\omega)=Y(\omega)-Y(\omega_L)$, where we subtract the fundamental harmonic yield due to the use of the logarithmic scale. 

\section{Results}

\subsection{Density Distribution}

\begin{figure}[h!]\centering
\includegraphics[width=0.5\textwidth]{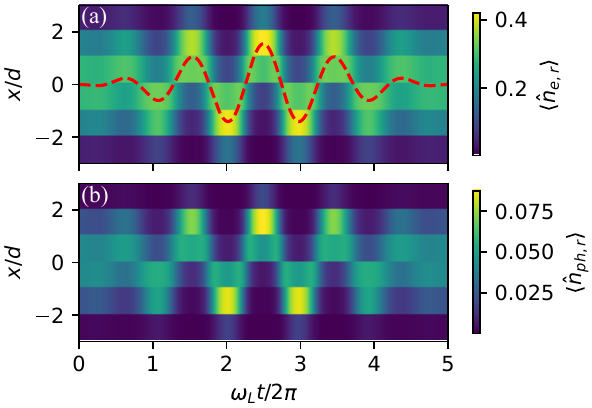}
	\caption[Simple model]{(a) The electron density distribution in Eq.\ref{eq:electrondensitydist} during the simulation. The dashed red line shows the electric field shape. (b) Phonon density distribution. We used the same coupling and phonon parameters as in Fig.~\ref{fig:energy-levels}, and the laser parameters were $\omega_L=0.002$, $n_{cyc}=5$, and $A_0=0.183$ (a.u). }\label{electron_phonon_density} 
\end{figure}

We begin by analyzing the electron and phonon clouds using the computed $a_n(t)$. Fig. \ref{electron_phonon_density} displays their distributions alongside the electric field profile (dashed red line). Panel (a) shows the electron density, calculated as:
\begin{align}\label{eq:electrondensitydist}
 \langle \hat n_{e,r}\rangle(t)= \sum_{m=0}^{\mathcal{N}_R-1}\sum_{n=0}^{\mathcal{N}_R-1} a_m^*(t) a_n(t) \bra{\phi_m} \hat n_{e,r} \ket{\phi_n}\ ,
\end{align}
with $0\le r\le 5 $. In the figure, we calculate the position of the electron and phonons at the $r$th site using $x=d(r-2.5)$. 
We observe in the figure that the electron cloud follows the shape of the electric field. 
Figure \ref{electron_phonon_density}.(b) shows the phonon density distribution, which we calculate using the previous equation, but substituting the electron number operator with the phonon number operator, $\langle \hat n_{ph,r}\rangle(t)$. 
We observe that the phonon density distribution follows the electron density distribution, which is a consequence of the Holstein interaction. 

\subsection{Harmonic spectrum}

\begin{figure}\centering
    \includegraphics[width=0.5\textwidth]{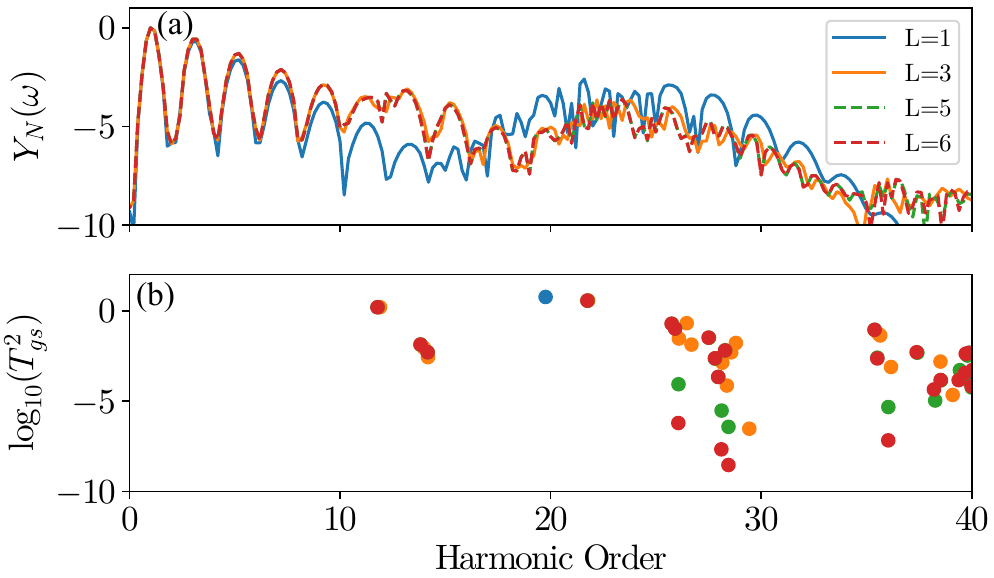}
    \caption{(a) The normalized harmonic yield $Y_N(\omega)$ as a function of the harmonic order for $L=\{1,3,5,6\}$. (b) The most relevant eigenstates based on $\log_{10}(T_\mathrm{gs}^2)$. The same coupling and laser parameters were used as in Fig.\ref{electron_phonon_density}.}
    \label{fig:hhg}
\end{figure}

Figure \ref{fig:hhg}.(a) shows the harmonic spectrum for the first 40 harmonic orders and $L=\{1,3,5,6\}$. 
For the case $L=1$, the Hilbert space has dimension $\mathcal{N}=6$, and we use $\mathcal{N}_R=6$ states in the simulation. For $L=3$, $\mathcal{N}=4374$ and we use $\mathcal{N}_R=1500$. For $L=5$, $\mathcal{N}=93750$ and we use $\mathcal N_R=1500$. As we increase $L$ the new states appear at higher energies, therefore $\mathcal N_R=1500$ is enough for $L=5$ and $L=6$. 
Figure \ref{fig:hhg}.(b) shows the eigenstates in function of $\log_{10}(T_\mathrm{gs}^2)$ and the harmonic order. 
For $L=1$, we observe the highest peak at the $1$st harmonic. The harmonic yield then decreases until around the $13$th harmonic. After the $13$th harmonic, the yield increases again, reaching a maximum around the $20$th harmonic. Beyond this point, the harmonic yield decreases. 
When observing the eigenstates for $L=1$ in Fig.\ref{fig:hhg}.(b), there is a state around the $20$th harmonic, this corresponds to the first excited state, which causes the increase in the harmonic yield around this region. 

We observe that for $L=3$ the results are converged because the yield data overlap with those for $L=5$ and $L=6$, which agrees with the ansatz stated before. 
For $L=\{3,5,6\}$, we see that around the $13$th harmonic order, the harmonics generated are around two orders of magnitude larger compared to $L=1$. This increase in the harmonics is due to new states that emerge from the e-ph interaction. Additionally, the inclusion of the e-ph coupling decreases the ground state energy, which also shifts the harmonic order of the excited states. This shift can be observed in Fig. \ref{fig:energy-levels} by comparing the harmonic order labels on the right. Moreover, this change in the harmonic order of the excited states also shifts the plateau of the harmonic yield around the $20$th harmonic. 

\begin{figure}
    \centering
    \includegraphics[width=1\linewidth]{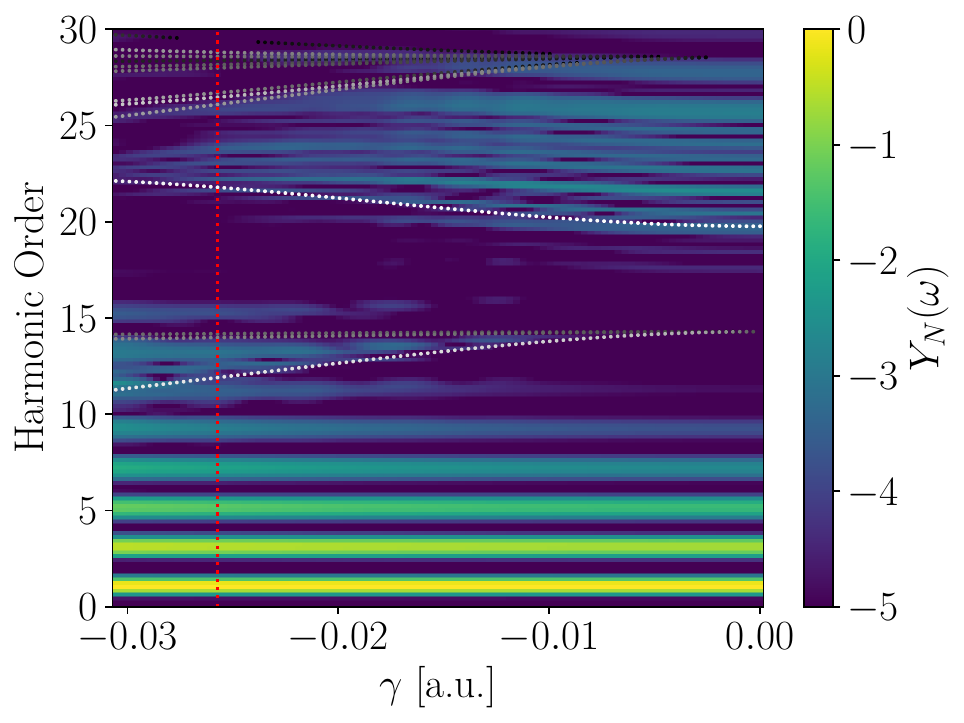}
    \caption{ $Y_N(\omega)$ as a function of the e-ph coupling $\gamma$. The points shown in gray scale correspond to states with a high $\log_{10}(T_\mathrm{gs}^2)$. The vertical red line indicates the $\gamma$ used in Fig.\ref{fig:hhg}.}
    \label{fig:Parameter_Scanning}
\end{figure}

In order to understand how different values of the e-ph interaction parameter $\gamma$ affect the harmonic yields, we plot the normalized harmonic yield $Y_N(\omega)$ for various $\gamma$ values in Fig. \ref{fig:Parameter_Scanning}. We also highlight with dots the magnitude order of the ground state transition vector (ranging from black to white according to $\log_{10}(T_\mathrm{gs}^2)$, where white indicates a high value and black a low value). 
The vertical dotted red line indicates the $\gamma$ value used for Fig.\ref{electron_phonon_density} and Fig.\ref{fig:hhg}. 
We observe that as $\gamma$ decreases from zero, the harmonic yield is enhanced around the $13$th harmonic, while the harmonics above the $20$th harmonics are diminished. 
This can be explained by the fact that as $\gamma$ decreases, the ground state transition value of the state near the $13$th harmonic changes from gray to white, indicating that this state becomes increasingly relevant. As this state gains relevance, the harmonic yield in this region increases, explaining the enhancement around $11$th and $15$th harmonics. 
Moreover, the decrease of the harmonic yield above the $20$th harmonic is also related to this state. As $\gamma$ decreases, the low energy states become dominant, and the probability amplitude concentrates more on these dominant states. This leads to a diminution in the population of the states above the $20$th harmonic, causing the harmonic yield in this region to diminish.  

\section{Conclusions}

In this work, we studied the effect of the Holstein polaron on HHG. 
From the Hamiltonian, we identified an e-ph binding energy by comparing the phononless case ($L=1$) with the case including phonons ($L>1$). 
From the simulation results, we examined the electron and phonon density distribution and observed that the electron cloud follows the shape of the electric laser field, as expected. Additionally, we found that the phonon cloud follows the movement of the electron cloud, indicating a binding between the electron and the phonons. 

The inclusion of phononic degrees of freedom introduces new energy levels accessible form the ground state can transition to during the simulation. These phonon induced states become increasingly relevant as the e-ph coupling strength $\gamma$ increases and $\omega_{ph}$ decreases. Simulations show that the harmonic yield is significantly modified by these new states, increasing by up to two orders of magnitude in regions where they emerge. As a result, emission from higher-energy states is reduced due to the dominant population of low-energy states, leading to a decreased harmonic yield near the first excited state of the phononless case. 

\section*{Acknowledgment}

We gratefully acknowledge funding by the German Research Foundation through 
IRTG 2676/1 ‘Imaging of Quantum Systems’, project no. 437567992, and 
CRC 1477 “Light-Matter Interactions at Interfaces,” Project
No. 441234705.

\section*{Data Availability}

All data generated and the code used in this study are available in GCA's Gitlab repository at https://gitlab.uni-rostock.de/zr2304/data-and-code-of-polaronic-effect-in-high-harmonic-generation.

\bibliography{main}

\begin{thebibliography}{66}%
\makeatletter
\providecommand \@ifxundefined [1]{%
 \@ifx{#1\undefined}
}%
\providecommand \@ifnum [1]{%
 \ifnum #1\expandafter \@firstoftwo
 \else \expandafter \@secondoftwo
 \fi
}%
\providecommand \@ifx [1]{%
 \ifx #1\expandafter \@firstoftwo
 \else \expandafter \@secondoftwo
 \fi
}%
\providecommand \natexlab [1]{#1}%
\providecommand \enquote  [1]{``#1''}%
\providecommand \bibnamefont  [1]{#1}%
\providecommand \bibfnamefont [1]{#1}%
\providecommand \citenamefont [1]{#1}%
\providecommand \href@noop [0]{\@secondoftwo}%
\providecommand \href [0]{\begingroup \@sanitize@url \@href}%
\providecommand \@href[1]{\@@startlink{#1}\@@href}%
\providecommand \@@href[1]{\endgroup#1\@@endlink}%
\providecommand \@sanitize@url [0]{\catcode `\\12\catcode `\$12\catcode
  `\&12\catcode `\#12\catcode `\^12\catcode `\_12\catcode `\%12\relax}%
\providecommand \@@startlink[1]{}%
\providecommand \@@endlink[0]{}%
\providecommand \url  [0]{\begingroup\@sanitize@url \@url }%
\providecommand \@url [1]{\endgroup\@href {#1}{\urlprefix }}%
\providecommand \urlprefix  [0]{URL }%
\providecommand \Eprint [0]{\href }%
\providecommand \doibase [0]{https://doi.org/}%
\providecommand \selectlanguage [0]{\@gobble}%
\providecommand \bibinfo  [0]{\@secondoftwo}%
\providecommand \bibfield  [0]{\@secondoftwo}%
\providecommand \translation [1]{[#1]}%
\providecommand \BibitemOpen [0]{}%
\providecommand \bibitemStop [0]{}%
\providecommand \bibitemNoStop [0]{.\EOS\space}%
\providecommand \EOS [0]{\spacefactor3000\relax}%
\providecommand \BibitemShut  [1]{\csname bibitem#1\endcsname}%
\let\auto@bib@innerbib\@empty
\bibitem [{\citenamefont {Goulielmakis}\ and\ \citenamefont
  {Brabec}(2022)}]{goulielmakis_high_2022}%
  \BibitemOpen
  \bibfield  {author} {\bibinfo {author} {\bibfnamefont {E.}~\bibnamefont
  {Goulielmakis}}\ and\ \bibinfo {author} {\bibfnamefont {T.}~\bibnamefont
  {Brabec}},\ }\bibfield  {title} {\bibinfo {title} {High harmonic generation
  in condensed matter},\ }\href {https://doi.org/10.1038/s41566-022-00988-y}
  {\bibfield  {journal} {\bibinfo  {journal} {Nature Photonics}\ }\textbf
  {\bibinfo {volume} {16}},\ \bibinfo {pages} {411} (\bibinfo {year}
  {2022})}\BibitemShut {NoStop}%
\bibitem [{\citenamefont {Lewenstein}\ \emph {et~al.}(1994)\citenamefont
  {Lewenstein}, \citenamefont {Balcou}, \citenamefont {Ivanov}, \citenamefont
  {L'Huillier},\ and\ \citenamefont {Corkum}}]{PhysRevA.49.2117}%
  \BibitemOpen
  \bibfield  {author} {\bibinfo {author} {\bibfnamefont {M.}~\bibnamefont
  {Lewenstein}}, \bibinfo {author} {\bibfnamefont {P.}~\bibnamefont {Balcou}},
  \bibinfo {author} {\bibfnamefont {M.~Y.}\ \bibnamefont {Ivanov}}, \bibinfo
  {author} {\bibfnamefont {A.}~\bibnamefont {L'Huillier}},\ and\ \bibinfo
  {author} {\bibfnamefont {P.~B.}\ \bibnamefont {Corkum}},\ }\bibfield  {title}
  {\bibinfo {title} {Theory of high-harmonic generation by low-frequency laser
  fields},\ }\href {https://doi.org/10.1103/PhysRevA.49.2117} {\bibfield
  {journal} {\bibinfo  {journal} {Phys. Rev. A}\ }\textbf {\bibinfo {volume}
  {49}},\ \bibinfo {pages} {2117} (\bibinfo {year} {1994})}\BibitemShut
  {NoStop}%
\bibitem [{\citenamefont {Corkum}(1993)}]{PhysRevLett.71.1994}%
  \BibitemOpen
  \bibfield  {author} {\bibinfo {author} {\bibfnamefont {P.~B.}\ \bibnamefont
  {Corkum}},\ }\bibfield  {title} {\bibinfo {title} {Plasma perspective on
  strong field multiphoton ionization},\ }\href
  {https://doi.org/10.1103/PhysRevLett.71.1994} {\bibfield  {journal} {\bibinfo
   {journal} {Phys. Rev. Lett.}\ }\textbf {\bibinfo {volume} {71}},\ \bibinfo
  {pages} {1994} (\bibinfo {year} {1993})}\BibitemShut {NoStop}%
\bibitem [{\citenamefont {Krausz}\ and\ \citenamefont
  {Ivanov}(2009)}]{attosecond}%
  \BibitemOpen
  \bibfield  {author} {\bibinfo {author} {\bibfnamefont {F.}~\bibnamefont
  {Krausz}}\ and\ \bibinfo {author} {\bibfnamefont {M.}~\bibnamefont
  {Ivanov}},\ }\bibfield  {title} {\bibinfo {title} {Attosecond physics},\
  }\href {https://doi.org/10.1103/RevModPhys.81.163} {\bibfield  {journal}
  {\bibinfo  {journal} {Rev. Mod. Phys.}\ }\textbf {\bibinfo {volume} {81}},\
  \bibinfo {pages} {163} (\bibinfo {year} {2009})}\BibitemShut {NoStop}%
\bibitem [{\citenamefont {Ghimire}\ \emph {et~al.}(2010)\citenamefont
  {Ghimire}, \citenamefont {DiChiara}, \citenamefont {Sistrunk}, \citenamefont
  {Agostini}, \citenamefont {DiMauro},\ and\ \citenamefont
  {Reis}}]{Ghimire2010}%
  \BibitemOpen
  \bibfield  {author} {\bibinfo {author} {\bibfnamefont {S.}~\bibnamefont
  {Ghimire}}, \bibinfo {author} {\bibfnamefont {A.~D.}\ \bibnamefont
  {DiChiara}}, \bibinfo {author} {\bibfnamefont {E.}~\bibnamefont {Sistrunk}},
  \bibinfo {author} {\bibfnamefont {P.}~\bibnamefont {Agostini}}, \bibinfo
  {author} {\bibfnamefont {L.~F.}\ \bibnamefont {DiMauro}},\ and\ \bibinfo
  {author} {\bibfnamefont {D.~A.}\ \bibnamefont {Reis}},\ }\bibfield  {title}
  {\bibinfo {title} {Observation of high-order harmonic generation in a bulk
  crystal},\ }\href {https://doi.org/10.1038/nphys1847} {\bibfield  {journal}
  {\bibinfo  {journal} {Nature Physics}\ }\textbf {\bibinfo {volume} {7}},\
  \bibinfo {pages} {138–141} (\bibinfo {year} {2010})}\BibitemShut {NoStop}%
\bibitem [{\citenamefont {Ghimire}\ and\ \citenamefont
  {Reis}(2019)}]{WOS:000454733100012}%
  \BibitemOpen
  \bibfield  {author} {\bibinfo {author} {\bibfnamefont {S.}~\bibnamefont
  {Ghimire}}\ and\ \bibinfo {author} {\bibfnamefont {D.~A.}\ \bibnamefont
  {Reis}},\ }\bibfield  {title} {\bibinfo {title} {High-harmonic generation
  from solids},\ }\href {https://doi.org/10.1038/s41567-018-0315-5} {\bibfield
  {journal} {\bibinfo  {journal} {NATURE PHYSICS}\ }\textbf {\bibinfo {volume}
  {15}},\ \bibinfo {pages} {10} (\bibinfo {year} {2019})}\BibitemShut {NoStop}%
\bibitem [{\citenamefont {Vampa}\ \emph {et~al.}(2015)\citenamefont {Vampa},
  \citenamefont {Hammond}, \citenamefont {Thir\'e}, \citenamefont {Schmidt},
  \citenamefont {L\'egar\'e}, \citenamefont {McDonald}, \citenamefont {Brabec},
  \citenamefont {Klug},\ and\ \citenamefont {Corkum}}]{structures}%
  \BibitemOpen
  \bibfield  {author} {\bibinfo {author} {\bibfnamefont {G.}~\bibnamefont
  {Vampa}}, \bibinfo {author} {\bibfnamefont {T.~J.}\ \bibnamefont {Hammond}},
  \bibinfo {author} {\bibfnamefont {N.}~\bibnamefont {Thir\'e}}, \bibinfo
  {author} {\bibfnamefont {B.~E.}\ \bibnamefont {Schmidt}}, \bibinfo {author}
  {\bibfnamefont {F.}~\bibnamefont {L\'egar\'e}}, \bibinfo {author}
  {\bibfnamefont {C.~R.}\ \bibnamefont {McDonald}}, \bibinfo {author}
  {\bibfnamefont {T.}~\bibnamefont {Brabec}}, \bibinfo {author} {\bibfnamefont
  {D.~D.}\ \bibnamefont {Klug}},\ and\ \bibinfo {author} {\bibfnamefont
  {P.~B.}\ \bibnamefont {Corkum}},\ }\bibfield  {title} {\bibinfo {title}
  {All-optical reconstruction of crystal band structure},\ }\href
  {https://doi.org/10.1103/PhysRevLett.115.193603} {\bibfield  {journal}
  {\bibinfo  {journal} {Phys. Rev. Lett.}\ }\textbf {\bibinfo {volume} {115}},\
  \bibinfo {pages} {193603} (\bibinfo {year} {2015})}\BibitemShut {NoStop}%
\bibitem [{\citenamefont {Luu}\ and\ \citenamefont {W\"{o}rner}(2018)}]{berry}%
  \BibitemOpen
  \bibfield  {author} {\bibinfo {author} {\bibfnamefont {T.~T.}\ \bibnamefont
  {Luu}}\ and\ \bibinfo {author} {\bibfnamefont {H.~J.}\ \bibnamefont
  {W\"{o}rner}},\ }\bibfield  {title} {\bibinfo {title} {Measurement of the
  berry curvature of solids using high-harmonic spectroscopy},\ }\bibfield
  {journal} {\bibinfo  {journal} {Nature Communications}\ }\textbf {\bibinfo
  {volume} {9}},\ \href {https://doi.org/10.1038/s41467-018-03397-4}
  {10.1038/s41467-018-03397-4} (\bibinfo {year} {2018})\BibitemShut {NoStop}%
\bibitem [{\citenamefont {Smirnova}\ \emph {et~al.}(2009)\citenamefont
  {Smirnova}, \citenamefont {Mairesse}, \citenamefont {Patchkovskii},
  \citenamefont {Dudovich}, \citenamefont {Villeneuve}, \citenamefont
  {Corkum},\ and\ \citenamefont {Ivanov}}]{smirnova_high_2009}%
  \BibitemOpen
  \bibfield  {author} {\bibinfo {author} {\bibfnamefont {O.}~\bibnamefont
  {Smirnova}}, \bibinfo {author} {\bibfnamefont {Y.}~\bibnamefont {Mairesse}},
  \bibinfo {author} {\bibfnamefont {S.}~\bibnamefont {Patchkovskii}}, \bibinfo
  {author} {\bibfnamefont {N.}~\bibnamefont {Dudovich}}, \bibinfo {author}
  {\bibfnamefont {D.}~\bibnamefont {Villeneuve}}, \bibinfo {author}
  {\bibfnamefont {P.}~\bibnamefont {Corkum}},\ and\ \bibinfo {author}
  {\bibfnamefont {M.~Y.}\ \bibnamefont {Ivanov}},\ }\bibfield  {title}
  {\bibinfo {title} {High harmonic interferometry of multi-electron dynamics in
  molecules},\ }\href {https://doi.org/10.1038/nature08253} {\bibfield
  {journal} {\bibinfo  {journal} {Nature}\ }\textbf {\bibinfo {volume} {460}},\
  \bibinfo {pages} {972} (\bibinfo {year} {2009})}\BibitemShut {NoStop}%
\bibitem [{\citenamefont {Bauer}\ and\ \citenamefont
  {Hansen}(2018)}]{BauerHansen2018}%
  \BibitemOpen
  \bibfield  {author} {\bibinfo {author} {\bibfnamefont {D.}~\bibnamefont
  {Bauer}}\ and\ \bibinfo {author} {\bibfnamefont {K.~K.}\ \bibnamefont
  {Hansen}},\ }\bibfield  {title} {\bibinfo {title} {High-harmonic generation
  in solids with and without topological edge states},\ }\href
  {https://doi.org/10.1103/PhysRevLett.120.177401} {\bibfield  {journal}
  {\bibinfo  {journal} {Phys. Rev. Lett.}\ }\textbf {\bibinfo {volume} {120}},\
  \bibinfo {pages} {177401} (\bibinfo {year} {2018})}\BibitemShut {NoStop}%
\bibitem [{\citenamefont {J\"ur\ss{}}\ and\ \citenamefont
  {Bauer}(2019)}]{JuerssBauer2019}%
  \BibitemOpen
  \bibfield  {author} {\bibinfo {author} {\bibfnamefont {H.}~\bibnamefont
  {J\"ur\ss{}}}\ and\ \bibinfo {author} {\bibfnamefont {D.}~\bibnamefont
  {Bauer}},\ }\bibfield  {title} {\bibinfo {title} {High-harmonic generation in
  su-schrieffer-heeger chains},\ }\href
  {https://doi.org/10.1103/PhysRevB.99.195428} {\bibfield  {journal} {\bibinfo
  {journal} {Phys. Rev. B}\ }\textbf {\bibinfo {volume} {99}},\ \bibinfo
  {pages} {195428} (\bibinfo {year} {2019})}\BibitemShut {NoStop}%
\bibitem [{\citenamefont {Pooyan}\ and\ \citenamefont
  {Bauer}(2025)}]{Pooyan_Bauer_2025}%
  \BibitemOpen
  \bibfield  {author} {\bibinfo {author} {\bibfnamefont {S.}~\bibnamefont
  {Pooyan}}\ and\ \bibinfo {author} {\bibfnamefont {D.}~\bibnamefont {Bauer}},\
  }\bibfield  {title} {\bibinfo {title} {Harmonic generation with topological
  edge states and electron-electron interaction},\ }\href
  {https://doi.org/10.1103/PhysRevB.111.014308} {\bibfield  {journal} {\bibinfo
   {journal} {Phys. Rev. B}\ }\textbf {\bibinfo {volume} {111}},\ \bibinfo
  {pages} {014308} (\bibinfo {year} {2025})}\BibitemShut {NoStop}%
\bibitem [{\citenamefont {Schubert}\ \emph {et~al.}(2014)\citenamefont
  {Schubert}, \citenamefont {Hohenleutner}, \citenamefont {Langer},
  \citenamefont {Urbanek}, \citenamefont {Lange}, \citenamefont {Huttner},
  \citenamefont {Golde}, \citenamefont {Meier}, \citenamefont {Kira},
  \citenamefont {Koch},\ and\ \citenamefont {Huber}}]{Schubert2014}%
  \BibitemOpen
  \bibfield  {author} {\bibinfo {author} {\bibfnamefont {O.}~\bibnamefont
  {Schubert}}, \bibinfo {author} {\bibfnamefont {M.}~\bibnamefont
  {Hohenleutner}}, \bibinfo {author} {\bibfnamefont {F.}~\bibnamefont
  {Langer}}, \bibinfo {author} {\bibfnamefont {B.}~\bibnamefont {Urbanek}},
  \bibinfo {author} {\bibfnamefont {C.}~\bibnamefont {Lange}}, \bibinfo
  {author} {\bibfnamefont {U.}~\bibnamefont {Huttner}}, \bibinfo {author}
  {\bibfnamefont {D.}~\bibnamefont {Golde}}, \bibinfo {author} {\bibfnamefont
  {T.}~\bibnamefont {Meier}}, \bibinfo {author} {\bibfnamefont
  {M.}~\bibnamefont {Kira}}, \bibinfo {author} {\bibfnamefont {S.~W.}\
  \bibnamefont {Koch}},\ and\ \bibinfo {author} {\bibfnamefont
  {R.}~\bibnamefont {Huber}},\ }\bibfield  {title} {\bibinfo {title} {Sub-cycle
  control of terahertz high-harmonic generation by dynamical bloch
  oscillations},\ }\href {https://doi.org/10.1038/nphoton.2013.349} {\bibfield
  {journal} {\bibinfo  {journal} {Nature Photonics}\ }\textbf {\bibinfo
  {volume} {8}},\ \bibinfo {pages} {119–123} (\bibinfo {year}
  {2014})}\BibitemShut {NoStop}%
\bibitem [{\citenamefont {Hohenleutner}\ \emph {et~al.}(2015)\citenamefont
  {Hohenleutner}, \citenamefont {Langer}, \citenamefont {Schubert},
  \citenamefont {Knorr}, \citenamefont {Huttner}, \citenamefont {Koch},
  \citenamefont {Kira},\ and\ \citenamefont {Huber}}]{Hohenleutner2015}%
  \BibitemOpen
  \bibfield  {author} {\bibinfo {author} {\bibfnamefont {M.}~\bibnamefont
  {Hohenleutner}}, \bibinfo {author} {\bibfnamefont {F.}~\bibnamefont
  {Langer}}, \bibinfo {author} {\bibfnamefont {O.}~\bibnamefont {Schubert}},
  \bibinfo {author} {\bibfnamefont {M.}~\bibnamefont {Knorr}}, \bibinfo
  {author} {\bibfnamefont {U.}~\bibnamefont {Huttner}}, \bibinfo {author}
  {\bibfnamefont {S.~W.}\ \bibnamefont {Koch}}, \bibinfo {author}
  {\bibfnamefont {M.}~\bibnamefont {Kira}},\ and\ \bibinfo {author}
  {\bibfnamefont {R.}~\bibnamefont {Huber}},\ }\bibfield  {title} {\bibinfo
  {title} {Real-time observation of interfering crystal electrons in
  high-harmonic generation},\ }\href {https://doi.org/10.1038/nature14652}
  {\bibfield  {journal} {\bibinfo  {journal} {Nature}\ }\textbf {\bibinfo
  {volume} {523}},\ \bibinfo {pages} {572–575} (\bibinfo {year}
  {2015})}\BibitemShut {NoStop}%
\bibitem [{\citenamefont {Luu}\ \emph {et~al.}(2015)\citenamefont {Luu},
  \citenamefont {Garg}, \citenamefont {Kruchinin}, \citenamefont {Moulet},
  \citenamefont {Hassan},\ and\ \citenamefont {Goulielmakis}}]{Luu2015}%
  \BibitemOpen
  \bibfield  {author} {\bibinfo {author} {\bibfnamefont {T.~T.}\ \bibnamefont
  {Luu}}, \bibinfo {author} {\bibfnamefont {M.}~\bibnamefont {Garg}}, \bibinfo
  {author} {\bibfnamefont {S.~Y.}\ \bibnamefont {Kruchinin}}, \bibinfo {author}
  {\bibfnamefont {A.}~\bibnamefont {Moulet}}, \bibinfo {author} {\bibfnamefont
  {M.~T.}\ \bibnamefont {Hassan}},\ and\ \bibinfo {author} {\bibfnamefont
  {E.}~\bibnamefont {Goulielmakis}},\ }\bibfield  {title} {\bibinfo {title}
  {Extreme ultraviolet high-harmonic spectroscopy of solids},\ }\href
  {https://doi.org/10.1038/nature14456} {\bibfield  {journal} {\bibinfo
  {journal} {Nature}\ }\textbf {\bibinfo {volume} {521}},\ \bibinfo {pages}
  {498–502} (\bibinfo {year} {2015})}\BibitemShut {NoStop}%
\bibitem [{\citenamefont {Ndabashimiye}\ \emph {et~al.}(2016)\citenamefont
  {Ndabashimiye}, \citenamefont {Ghimire}, \citenamefont {Wu}, \citenamefont
  {Browne}, \citenamefont {Schafer}, \citenamefont {Gaarde},\ and\
  \citenamefont {Reis}}]{Ndabashimiye2016}%
  \BibitemOpen
  \bibfield  {author} {\bibinfo {author} {\bibfnamefont {G.}~\bibnamefont
  {Ndabashimiye}}, \bibinfo {author} {\bibfnamefont {S.}~\bibnamefont
  {Ghimire}}, \bibinfo {author} {\bibfnamefont {M.}~\bibnamefont {Wu}},
  \bibinfo {author} {\bibfnamefont {D.~A.}\ \bibnamefont {Browne}}, \bibinfo
  {author} {\bibfnamefont {K.~J.}\ \bibnamefont {Schafer}}, \bibinfo {author}
  {\bibfnamefont {M.~B.}\ \bibnamefont {Gaarde}},\ and\ \bibinfo {author}
  {\bibfnamefont {D.~A.}\ \bibnamefont {Reis}},\ }\bibfield  {title} {\bibinfo
  {title} {Solid-state harmonics beyond the atomic limit},\ }\href
  {https://doi.org/10.1038/nature17660} {\bibfield  {journal} {\bibinfo
  {journal} {Nature}\ }\textbf {\bibinfo {volume} {534}},\ \bibinfo {pages}
  {520–523} (\bibinfo {year} {2016})}\BibitemShut {NoStop}%
\bibitem [{\citenamefont {Langer}\ \emph {et~al.}(2017)\citenamefont {Langer},
  \citenamefont {Hohenleutner}, \citenamefont {Huttner}, \citenamefont {Koch},
  \citenamefont {Kira},\ and\ \citenamefont {Huber}}]{Langer2017}%
  \BibitemOpen
  \bibfield  {author} {\bibinfo {author} {\bibfnamefont {F.}~\bibnamefont
  {Langer}}, \bibinfo {author} {\bibfnamefont {M.}~\bibnamefont
  {Hohenleutner}}, \bibinfo {author} {\bibfnamefont {U.}~\bibnamefont
  {Huttner}}, \bibinfo {author} {\bibfnamefont {S.~W.}\ \bibnamefont {Koch}},
  \bibinfo {author} {\bibfnamefont {M.}~\bibnamefont {Kira}},\ and\ \bibinfo
  {author} {\bibfnamefont {R.}~\bibnamefont {Huber}},\ }\bibfield  {title}
  {\bibinfo {title} {Symmetry-controlled temporal structure of high-harmonic
  carrier fields from a bulk crystal},\ }\href
  {https://doi.org/10.1038/nphoton.2017.29} {\bibfield  {journal} {\bibinfo
  {journal} {Nature Photonics}\ }\textbf {\bibinfo {volume} {11}},\ \bibinfo
  {pages} {227–231} (\bibinfo {year} {2017})}\BibitemShut {NoStop}%
\bibitem [{\citenamefont {Tancogne-Dejean}\ \emph {et~al.}(2017)\citenamefont
  {Tancogne-Dejean}, \citenamefont {M\"ucke}, \citenamefont {K\"artner},\ and\
  \citenamefont {Rubio}}]{PhysRevLett.118.087403}%
  \BibitemOpen
  \bibfield  {author} {\bibinfo {author} {\bibfnamefont {N.}~\bibnamefont
  {Tancogne-Dejean}}, \bibinfo {author} {\bibfnamefont {O.~D.}\ \bibnamefont
  {M\"ucke}}, \bibinfo {author} {\bibfnamefont {F.~X.}\ \bibnamefont
  {K\"artner}},\ and\ \bibinfo {author} {\bibfnamefont {A.}~\bibnamefont
  {Rubio}},\ }\bibfield  {title} {\bibinfo {title} {Impact of the electronic
  band structure in high-harmonic generation spectra of solids},\ }\href
  {https://doi.org/10.1103/PhysRevLett.118.087403} {\bibfield  {journal}
  {\bibinfo  {journal} {Phys. Rev. Lett.}\ }\textbf {\bibinfo {volume} {118}},\
  \bibinfo {pages} {087403} (\bibinfo {year} {2017})}\BibitemShut {NoStop}%
\bibitem [{\citenamefont {You}\ \emph {et~al.}(2017)\citenamefont {You},
  \citenamefont {Yin}, \citenamefont {Wu}, \citenamefont {Chew}, \citenamefont
  {Ren}, \citenamefont {Zhuang}, \citenamefont {Gholam-Mirzaei}, \citenamefont
  {Chini}, \citenamefont {Chang},\ and\ \citenamefont {Ghimire}}]{You2017}%
  \BibitemOpen
  \bibfield  {author} {\bibinfo {author} {\bibfnamefont {Y.~S.}\ \bibnamefont
  {You}}, \bibinfo {author} {\bibfnamefont {Y.}~\bibnamefont {Yin}}, \bibinfo
  {author} {\bibfnamefont {Y.}~\bibnamefont {Wu}}, \bibinfo {author}
  {\bibfnamefont {A.}~\bibnamefont {Chew}}, \bibinfo {author} {\bibfnamefont
  {X.}~\bibnamefont {Ren}}, \bibinfo {author} {\bibfnamefont {F.}~\bibnamefont
  {Zhuang}}, \bibinfo {author} {\bibfnamefont {S.}~\bibnamefont
  {Gholam-Mirzaei}}, \bibinfo {author} {\bibfnamefont {M.}~\bibnamefont
  {Chini}}, \bibinfo {author} {\bibfnamefont {Z.}~\bibnamefont {Chang}},\ and\
  \bibinfo {author} {\bibfnamefont {S.}~\bibnamefont {Ghimire}},\ }\bibfield
  {title} {\bibinfo {title} {High-harmonic generation in amorphous solids},\
  }\bibfield  {journal} {\bibinfo  {journal} {Nature Communications}\ }\textbf
  {\bibinfo {volume} {8}},\ \href {https://doi.org/10.1038/s41467-017-00989-4}
  {10.1038/s41467-017-00989-4} (\bibinfo {year} {2017})\BibitemShut {NoStop}%
\bibitem [{\citenamefont {Zhang}\ \emph {et~al.}(2018)\citenamefont {Zhang},
  \citenamefont {Si}, \citenamefont {Murakami}, \citenamefont {Bai},\ and\
  \citenamefont {George}}]{Zhang2018}%
  \BibitemOpen
  \bibfield  {author} {\bibinfo {author} {\bibfnamefont {G.~P.}\ \bibnamefont
  {Zhang}}, \bibinfo {author} {\bibfnamefont {M.~S.}\ \bibnamefont {Si}},
  \bibinfo {author} {\bibfnamefont {M.}~\bibnamefont {Murakami}}, \bibinfo
  {author} {\bibfnamefont {Y.~H.}\ \bibnamefont {Bai}},\ and\ \bibinfo {author}
  {\bibfnamefont {T.~F.}\ \bibnamefont {George}},\ }\bibfield  {title}
  {\bibinfo {title} {Generating high-order optical and spin harmonics from
  ferromagnetic monolayers},\ }\bibfield  {journal} {\bibinfo  {journal}
  {Nature Communications}\ }\textbf {\bibinfo {volume} {9}},\ \href
  {https://doi.org/10.1038/s41467-018-05535-4} {10.1038/s41467-018-05535-4}
  (\bibinfo {year} {2018})\BibitemShut {NoStop}%
\bibitem [{\citenamefont {Vampa}\ \emph {et~al.}(2018)\citenamefont {Vampa},
  \citenamefont {Hammond}, \citenamefont {Taucer}, \citenamefont {Ding},
  \citenamefont {Ropagnol}, \citenamefont {Ozaki}, \citenamefont {Delprat},
  \citenamefont {Chaker}, \citenamefont {Thiré}, \citenamefont {Schmidt},
  \citenamefont {Légaré}, \citenamefont {Klug}, \citenamefont {Naumov},
  \citenamefont {Villeneuve}, \citenamefont {Staudte},\ and\ \citenamefont
  {Corkum}}]{Vampa2018}%
  \BibitemOpen
  \bibfield  {author} {\bibinfo {author} {\bibfnamefont {G.}~\bibnamefont
  {Vampa}}, \bibinfo {author} {\bibfnamefont {T.~J.}\ \bibnamefont {Hammond}},
  \bibinfo {author} {\bibfnamefont {M.}~\bibnamefont {Taucer}}, \bibinfo
  {author} {\bibfnamefont {X.}~\bibnamefont {Ding}}, \bibinfo {author}
  {\bibfnamefont {X.}~\bibnamefont {Ropagnol}}, \bibinfo {author}
  {\bibfnamefont {T.}~\bibnamefont {Ozaki}}, \bibinfo {author} {\bibfnamefont
  {S.}~\bibnamefont {Delprat}}, \bibinfo {author} {\bibfnamefont
  {M.}~\bibnamefont {Chaker}}, \bibinfo {author} {\bibfnamefont
  {N.}~\bibnamefont {Thiré}}, \bibinfo {author} {\bibfnamefont {B.~E.}\
  \bibnamefont {Schmidt}}, \bibinfo {author} {\bibfnamefont {F.}~\bibnamefont
  {Légaré}}, \bibinfo {author} {\bibfnamefont {D.~D.}\ \bibnamefont {Klug}},
  \bibinfo {author} {\bibfnamefont {A.~Y.}\ \bibnamefont {Naumov}}, \bibinfo
  {author} {\bibfnamefont {D.~M.}\ \bibnamefont {Villeneuve}}, \bibinfo
  {author} {\bibfnamefont {A.}~\bibnamefont {Staudte}},\ and\ \bibinfo {author}
  {\bibfnamefont {P.~B.}\ \bibnamefont {Corkum}},\ }\bibfield  {title}
  {\bibinfo {title} {Strong-field optoelectronics in solids},\ }\href
  {https://doi.org/10.1038/s41566-018-0193-5} {\bibfield  {journal} {\bibinfo
  {journal} {Nature Photonics}\ }\textbf {\bibinfo {volume} {12}},\ \bibinfo
  {pages} {465–468} (\bibinfo {year} {2018})}\BibitemShut {NoStop}%
\bibitem [{\citenamefont {Baudisch}\ \emph {et~al.}(2018)\citenamefont
  {Baudisch}, \citenamefont {Marini}, \citenamefont {Cox}, \citenamefont {Zhu},
  \citenamefont {Silva}, \citenamefont {Teichmann}, \citenamefont {Massicotte},
  \citenamefont {Koppens}, \citenamefont {Levitov}, \citenamefont {García~de
  Abajo},\ and\ \citenamefont {Biegert}}]{Baudisch2018}%
  \BibitemOpen
  \bibfield  {author} {\bibinfo {author} {\bibfnamefont {M.}~\bibnamefont
  {Baudisch}}, \bibinfo {author} {\bibfnamefont {A.}~\bibnamefont {Marini}},
  \bibinfo {author} {\bibfnamefont {J.~D.}\ \bibnamefont {Cox}}, \bibinfo
  {author} {\bibfnamefont {T.}~\bibnamefont {Zhu}}, \bibinfo {author}
  {\bibfnamefont {F.}~\bibnamefont {Silva}}, \bibinfo {author} {\bibfnamefont
  {S.}~\bibnamefont {Teichmann}}, \bibinfo {author} {\bibfnamefont
  {M.}~\bibnamefont {Massicotte}}, \bibinfo {author} {\bibfnamefont
  {F.}~\bibnamefont {Koppens}}, \bibinfo {author} {\bibfnamefont {L.~S.}\
  \bibnamefont {Levitov}}, \bibinfo {author} {\bibfnamefont {F.~J.}\
  \bibnamefont {García~de Abajo}},\ and\ \bibinfo {author} {\bibfnamefont
  {J.}~\bibnamefont {Biegert}},\ }\bibfield  {title} {\bibinfo {title}
  {Ultrafast nonlinear optical response of dirac fermions in graphene},\
  }\bibfield  {journal} {\bibinfo  {journal} {Nature Communications}\ }\textbf
  {\bibinfo {volume} {9}},\ \href {https://doi.org/10.1038/s41467-018-03413-7}
  {10.1038/s41467-018-03413-7} (\bibinfo {year} {2018})\BibitemShut {NoStop}%
\bibitem [{\citenamefont {Garg}\ \emph {et~al.}(2018)\citenamefont {Garg},
  \citenamefont {Kim},\ and\ \citenamefont {Goulielmakis}}]{Garg2018}%
  \BibitemOpen
  \bibfield  {author} {\bibinfo {author} {\bibfnamefont {M.}~\bibnamefont
  {Garg}}, \bibinfo {author} {\bibfnamefont {H.~Y.}\ \bibnamefont {Kim}},\ and\
  \bibinfo {author} {\bibfnamefont {E.}~\bibnamefont {Goulielmakis}},\
  }\bibfield  {title} {\bibinfo {title} {Ultimate waveform reproducibility of
  extreme-ultraviolet pulses by high-harmonic generation in quartz},\ }\href
  {https://doi.org/10.1038/s41566-018-0123-6} {\bibfield  {journal} {\bibinfo
  {journal} {Nature Photonics}\ }\textbf {\bibinfo {volume} {12}},\ \bibinfo
  {pages} {291–296} (\bibinfo {year} {2018})}\BibitemShut {NoStop}%
\bibitem [{\citenamefont {Dr\"ueke}\ and\ \citenamefont
  {Bauer}(2019)}]{PhysRevA.99.053402}%
  \BibitemOpen
  \bibfield  {author} {\bibinfo {author} {\bibfnamefont {H.}~\bibnamefont
  {Dr\"ueke}}\ and\ \bibinfo {author} {\bibfnamefont {D.}~\bibnamefont
  {Bauer}},\ }\bibfield  {title} {\bibinfo {title} {Robustness of topologically
  sensitive harmonic generation in laser-driven linear chains},\ }\href
  {https://doi.org/10.1103/PhysRevA.99.053402} {\bibfield  {journal} {\bibinfo
  {journal} {Phys. Rev. A}\ }\textbf {\bibinfo {volume} {99}},\ \bibinfo
  {pages} {053402} (\bibinfo {year} {2019})}\BibitemShut {NoStop}%
\bibitem [{\citenamefont {Moos}\ \emph {et~al.}(2020)\citenamefont {Moos},
  \citenamefont {J\"ur\ss{}},\ and\ \citenamefont
  {Bauer}}]{PhysRevA.102.053112}%
  \BibitemOpen
  \bibfield  {author} {\bibinfo {author} {\bibfnamefont {D.}~\bibnamefont
  {Moos}}, \bibinfo {author} {\bibfnamefont {H.}~\bibnamefont {J\"ur\ss{}}},\
  and\ \bibinfo {author} {\bibfnamefont {D.}~\bibnamefont {Bauer}},\ }\bibfield
   {title} {\bibinfo {title} {Intense-laser-driven electron dynamics and
  high-order harmonic generation in solids including topological effects},\
  }\href {https://doi.org/10.1103/PhysRevA.102.053112} {\bibfield  {journal}
  {\bibinfo  {journal} {Phys. Rev. A}\ }\textbf {\bibinfo {volume} {102}},\
  \bibinfo {pages} {053112} (\bibinfo {year} {2020})}\BibitemShut {NoStop}%
\bibitem [{\citenamefont {Koochaki~Kelardeh}\ \emph {et~al.}(2017)\citenamefont
  {Koochaki~Kelardeh}, \citenamefont {Apalkov},\ and\ \citenamefont
  {Stockman}}]{PhysRevB.96.075409}%
  \BibitemOpen
  \bibfield  {author} {\bibinfo {author} {\bibfnamefont {H.}~\bibnamefont
  {Koochaki~Kelardeh}}, \bibinfo {author} {\bibfnamefont {V.}~\bibnamefont
  {Apalkov}},\ and\ \bibinfo {author} {\bibfnamefont {M.~I.}\ \bibnamefont
  {Stockman}},\ }\bibfield  {title} {\bibinfo {title} {Graphene superlattices
  in strong circularly polarized fields: Chirality, berry phase, and attosecond
  dynamics},\ }\href {https://doi.org/10.1103/PhysRevB.96.075409} {\bibfield
  {journal} {\bibinfo  {journal} {Phys. Rev. B}\ }\textbf {\bibinfo {volume}
  {96}},\ \bibinfo {pages} {075409} (\bibinfo {year} {2017})}\BibitemShut
  {NoStop}%
\bibitem [{\citenamefont {Neufeld}\ \emph {et~al.}(2023)\citenamefont
  {Neufeld}, \citenamefont {Tancogne-Dejean}, \citenamefont {H\"ubener},
  \citenamefont {De~Giovannini},\ and\ \citenamefont
  {Rubio}}]{PhysRevX.13.031011}%
  \BibitemOpen
  \bibfield  {author} {\bibinfo {author} {\bibfnamefont {O.}~\bibnamefont
  {Neufeld}}, \bibinfo {author} {\bibfnamefont {N.}~\bibnamefont
  {Tancogne-Dejean}}, \bibinfo {author} {\bibfnamefont {H.}~\bibnamefont
  {H\"ubener}}, \bibinfo {author} {\bibfnamefont {U.}~\bibnamefont
  {De~Giovannini}},\ and\ \bibinfo {author} {\bibfnamefont {A.}~\bibnamefont
  {Rubio}},\ }\bibfield  {title} {\bibinfo {title} {Are there universal
  signatures of topological phases in high-harmonic generation? probably
  not.},\ }\href {https://doi.org/10.1103/PhysRevX.13.031011} {\bibfield
  {journal} {\bibinfo  {journal} {Phys. Rev. X}\ }\textbf {\bibinfo {volume}
  {13}},\ \bibinfo {pages} {031011} (\bibinfo {year} {2023})}\BibitemShut
  {NoStop}%
\bibitem [{\citenamefont {Luu}\ and\ \citenamefont
  {Wörner}(2018)}]{luu_measurement_2018}%
  \BibitemOpen
  \bibfield  {author} {\bibinfo {author} {\bibfnamefont {T.~T.}\ \bibnamefont
  {Luu}}\ and\ \bibinfo {author} {\bibfnamefont {H.~J.}\ \bibnamefont
  {Wörner}},\ }\bibfield  {title} {\bibinfo {title} {Measurement of the
  {Berry} curvature of solids using high-harmonic spectroscopy},\ }\href
  {https://doi.org/10.1038/s41467-018-03397-4} {\bibfield  {journal} {\bibinfo
  {journal} {Nature Communications}\ }\textbf {\bibinfo {volume} {9}},\
  \bibinfo {pages} {916} (\bibinfo {year} {2018})}\BibitemShut {NoStop}%
\bibitem [{\citenamefont {Reimann}\ \emph {et~al.}(2018)\citenamefont
  {Reimann}, \citenamefont {Schlauderer}, \citenamefont {Schmid}, \citenamefont
  {Langer}, \citenamefont {Baierl}, \citenamefont {Kokh}, \citenamefont
  {Tereshchenko}, \citenamefont {Kimura}, \citenamefont {Lange}, \citenamefont
  {Güdde}, \citenamefont {Höfer},\ and\ \citenamefont
  {Huber}}]{reimann_subcycle_2018}%
  \BibitemOpen
  \bibfield  {author} {\bibinfo {author} {\bibfnamefont {J.}~\bibnamefont
  {Reimann}}, \bibinfo {author} {\bibfnamefont {S.}~\bibnamefont
  {Schlauderer}}, \bibinfo {author} {\bibfnamefont {C.~P.}\ \bibnamefont
  {Schmid}}, \bibinfo {author} {\bibfnamefont {F.}~\bibnamefont {Langer}},
  \bibinfo {author} {\bibfnamefont {S.}~\bibnamefont {Baierl}}, \bibinfo
  {author} {\bibfnamefont {K.~A.}\ \bibnamefont {Kokh}}, \bibinfo {author}
  {\bibfnamefont {O.~E.}\ \bibnamefont {Tereshchenko}}, \bibinfo {author}
  {\bibfnamefont {A.}~\bibnamefont {Kimura}}, \bibinfo {author} {\bibfnamefont
  {C.}~\bibnamefont {Lange}}, \bibinfo {author} {\bibfnamefont
  {J.}~\bibnamefont {Güdde}}, \bibinfo {author} {\bibfnamefont
  {U.}~\bibnamefont {Höfer}},\ and\ \bibinfo {author} {\bibfnamefont
  {R.}~\bibnamefont {Huber}},\ }\bibfield  {title} {\bibinfo {title} {Subcycle
  observation of lightwave-driven {Dirac} currents in a topological surface
  band},\ }\href {https://doi.org/10.1038/s41586-018-0544-x} {\bibfield
  {journal} {\bibinfo  {journal} {Nature}\ }\textbf {\bibinfo {volume} {562}},\
  \bibinfo {pages} {396} (\bibinfo {year} {2018})}\BibitemShut {NoStop}%
\bibitem [{\citenamefont {Korobenko}\ \emph {et~al.}(2021)\citenamefont
  {Korobenko}, \citenamefont {Saha}, \citenamefont {Godfrey}, \citenamefont
  {Gertsvolf}, \citenamefont {Naumov}, \citenamefont {Villeneuve},
  \citenamefont {Boltasseva}, \citenamefont {Shalaev},\ and\ \citenamefont
  {Corkum}}]{korobenko_high-harmonic_2021}%
  \BibitemOpen
  \bibfield  {author} {\bibinfo {author} {\bibfnamefont {A.}~\bibnamefont
  {Korobenko}}, \bibinfo {author} {\bibfnamefont {S.}~\bibnamefont {Saha}},
  \bibinfo {author} {\bibfnamefont {A.~T.~K.}\ \bibnamefont {Godfrey}},
  \bibinfo {author} {\bibfnamefont {M.}~\bibnamefont {Gertsvolf}}, \bibinfo
  {author} {\bibfnamefont {A.~Y.}\ \bibnamefont {Naumov}}, \bibinfo {author}
  {\bibfnamefont {D.~M.}\ \bibnamefont {Villeneuve}}, \bibinfo {author}
  {\bibfnamefont {A.}~\bibnamefont {Boltasseva}}, \bibinfo {author}
  {\bibfnamefont {V.~M.}\ \bibnamefont {Shalaev}},\ and\ \bibinfo {author}
  {\bibfnamefont {P.~B.}\ \bibnamefont {Corkum}},\ }\bibfield  {title}
  {\bibinfo {title} {High-harmonic generation in metallic titanium nitride},\
  }\href {https://doi.org/10.1038/s41467-021-25224-z} {\bibfield  {journal}
  {\bibinfo  {journal} {Nature Communications}\ }\textbf {\bibinfo {volume}
  {12}},\ \bibinfo {pages} {4981} (\bibinfo {year} {2021})}\BibitemShut
  {NoStop}%
\bibitem [{\citenamefont {Li}\ \emph {et~al.}(2023)\citenamefont {Li},
  \citenamefont {Saleh}, \citenamefont {Sharma}, \citenamefont {Hünecke},
  \citenamefont {Sierka}, \citenamefont {Neuhaus}, \citenamefont {Hedewig},
  \citenamefont {Bergues}, \citenamefont {Alharbi}, \citenamefont {ALQahtani},
  \citenamefont {Azzeer}, \citenamefont {Gräfe}, \citenamefont {Kling},
  \citenamefont {Alharbi},\ and\ \citenamefont
  {Wang}}]{https://doi.org/10.1002/adom.202203070}%
  \BibitemOpen
  \bibfield  {author} {\bibinfo {author} {\bibfnamefont {W.}~\bibnamefont
  {Li}}, \bibinfo {author} {\bibfnamefont {A.}~\bibnamefont {Saleh}}, \bibinfo
  {author} {\bibfnamefont {M.}~\bibnamefont {Sharma}}, \bibinfo {author}
  {\bibfnamefont {C.}~\bibnamefont {Hünecke}}, \bibinfo {author}
  {\bibfnamefont {M.}~\bibnamefont {Sierka}}, \bibinfo {author} {\bibfnamefont
  {M.}~\bibnamefont {Neuhaus}}, \bibinfo {author} {\bibfnamefont
  {L.}~\bibnamefont {Hedewig}}, \bibinfo {author} {\bibfnamefont
  {B.}~\bibnamefont {Bergues}}, \bibinfo {author} {\bibfnamefont
  {M.}~\bibnamefont {Alharbi}}, \bibinfo {author} {\bibfnamefont
  {H.}~\bibnamefont {ALQahtani}}, \bibinfo {author} {\bibfnamefont {A.~M.}\
  \bibnamefont {Azzeer}}, \bibinfo {author} {\bibfnamefont {S.}~\bibnamefont
  {Gräfe}}, \bibinfo {author} {\bibfnamefont {M.~F.}\ \bibnamefont {Kling}},
  \bibinfo {author} {\bibfnamefont {A.~F.}\ \bibnamefont {Alharbi}},\ and\
  \bibinfo {author} {\bibfnamefont {Z.}~\bibnamefont {Wang}},\ }\bibfield
  {title} {\bibinfo {title} {Resonance effect in brunel harmonic generation in
  thin film organic semiconductors},\ }\href
  {https://doi.org/https://doi.org/10.1002/adom.202203070} {\bibfield
  {journal} {\bibinfo  {journal} {Advanced Optical Materials}\ }\textbf
  {\bibinfo {volume} {11}},\ \bibinfo {pages} {2203070} (\bibinfo {year}
  {2023})},\ \Eprint
  {https://arxiv.org/abs/https://advanced.onlinelibrary.wiley.com/doi/pdf/10.1002/adom.202203070}
  {https://advanced.onlinelibrary.wiley.com/doi/pdf/10.1002/adom.202203070}
  \BibitemShut {NoStop}%
\bibitem [{\citenamefont {Annunziata}\ \emph {et~al.}(2024)\citenamefont
  {Annunziata}, \citenamefont {Becker}, \citenamefont {Murillo-Sánchez},
  \citenamefont {Friebel}, \citenamefont {Stagira}, \citenamefont {Faccialà},
  \citenamefont {Vozzi},\ and\ \citenamefont {Cattaneo}}]{10.1063/5.0191184}%
  \BibitemOpen
  \bibfield  {author} {\bibinfo {author} {\bibfnamefont {A.}~\bibnamefont
  {Annunziata}}, \bibinfo {author} {\bibfnamefont {L.}~\bibnamefont {Becker}},
  \bibinfo {author} {\bibfnamefont {M.~L.}\ \bibnamefont {Murillo-Sánchez}},
  \bibinfo {author} {\bibfnamefont {P.}~\bibnamefont {Friebel}}, \bibinfo
  {author} {\bibfnamefont {S.}~\bibnamefont {Stagira}}, \bibinfo {author}
  {\bibfnamefont {D.}~\bibnamefont {Faccialà}}, \bibinfo {author}
  {\bibfnamefont {C.}~\bibnamefont {Vozzi}},\ and\ \bibinfo {author}
  {\bibfnamefont {L.}~\bibnamefont {Cattaneo}},\ }\bibfield  {title} {\bibinfo
  {title} {High-order harmonic generation in liquid crystals},\ }\href
  {https://doi.org/10.1063/5.0191184} {\bibfield  {journal} {\bibinfo
  {journal} {APL Photonics}\ }\textbf {\bibinfo {volume} {9}},\ \bibinfo
  {pages} {060801} (\bibinfo {year} {2024})},\ \Eprint
  {https://arxiv.org/abs/https://pubs.aip.org/aip/app/article-pdf/doi/10.1063/5.0191184/20038545/060801\_1\_5.0191184.pdf}
  {https://pubs.aip.org/aip/app/article-pdf/doi/10.1063/5.0191184/20038545/060801\_1\_5.0191184.pdf}
  \BibitemShut {NoStop}%
\bibitem [{\citenamefont {Venkatesh}\ \emph {et~al.}(2023)\citenamefont
  {Venkatesh}, \citenamefont {Kim}, \citenamefont {Boltaev}, \citenamefont
  {Konda}, \citenamefont {Svedlindh}, \citenamefont {Li},\ and\ \citenamefont
  {Ganeev}}]{WOS:000969760700001}%
  \BibitemOpen
  \bibfield  {author} {\bibinfo {author} {\bibfnamefont {M.}~\bibnamefont
  {Venkatesh}}, \bibinfo {author} {\bibfnamefont {V.~V.~V.}\ \bibnamefont
  {Kim}}, \bibinfo {author} {\bibfnamefont {G.~S.~S.}\ \bibnamefont {Boltaev}},
  \bibinfo {author} {\bibfnamefont {S.~R.}\ \bibnamefont {Konda}}, \bibinfo
  {author} {\bibfnamefont {P.}~\bibnamefont {Svedlindh}}, \bibinfo {author}
  {\bibfnamefont {W.}~\bibnamefont {Li}},\ and\ \bibinfo {author}
  {\bibfnamefont {R.~A.~A.}\ \bibnamefont {Ganeev}},\ }\bibfield  {title}
  {\bibinfo {title} {High-order harmonics generation in mos2 transition metal
  dichalcogenides: Effect of nickel and carbon nanotube dopants},\ }\bibfield
  {journal} {\bibinfo  {journal} {INTERNATIONAL JOURNAL OF MOLECULAR SCIENCES}\
  }\textbf {\bibinfo {volume} {24}},\ \href
  {https://doi.org/10.3390/ijms24076540} {10.3390/ijms24076540} (\bibinfo
  {year} {2023})\BibitemShut {NoStop}%
\bibitem [{\citenamefont {B\^aldea}\ \emph {et~al.}(2004)\citenamefont
  {B\^aldea}, \citenamefont {Gupta}, \citenamefont {Cederbaum},\ and\
  \citenamefont {Moiseyev}}]{PhysRevB.69.245311}%
  \BibitemOpen
  \bibfield  {author} {\bibinfo {author} {\bibfnamefont {I.}~\bibnamefont
  {B\^aldea}}, \bibinfo {author} {\bibfnamefont {A.~K.}\ \bibnamefont {Gupta}},
  \bibinfo {author} {\bibfnamefont {L.~S.}\ \bibnamefont {Cederbaum}},\ and\
  \bibinfo {author} {\bibfnamefont {N.}~\bibnamefont {Moiseyev}},\ }\bibfield
  {title} {\bibinfo {title} {High-harmonic generation by quantum-dot
  nanorings},\ }\href {https://doi.org/10.1103/PhysRevB.69.245311} {\bibfield
  {journal} {\bibinfo  {journal} {Phys. Rev. B}\ }\textbf {\bibinfo {volume}
  {69}},\ \bibinfo {pages} {245311} (\bibinfo {year} {2004})}\BibitemShut
  {NoStop}%
\bibitem [{\citenamefont {Austin}\ and\ \citenamefont
  {and}(2001)}]{Austin01112001}%
  \BibitemOpen
  \bibfield  {author} {\bibinfo {author} {\bibfnamefont {I.~G.}\ \bibnamefont
  {Austin}}\ and\ \bibinfo {author} {\bibfnamefont {N.~F.~M.}\ \bibnamefont
  {and}},\ }\bibfield  {title} {\bibinfo {title} {Polarons in crystalline and
  non-crystalline materials},\ }\href
  {https://doi.org/10.1080/00018730110103249} {\bibfield  {journal} {\bibinfo
  {journal} {Advances in Physics}\ }\textbf {\bibinfo {volume} {50}},\ \bibinfo
  {pages} {757} (\bibinfo {year} {2001})},\ \Eprint
  {https://arxiv.org/abs/https://doi.org/10.1080/00018730110103249}
  {https://doi.org/10.1080/00018730110103249} \BibitemShut {NoStop}%
\bibitem [{\citenamefont {Bogomolov}\ and\ \citenamefont
  {Mirlin}(1968)}]{https://doi.org/10.1002/pssb.19680270144}%
  \BibitemOpen
  \bibfield  {author} {\bibinfo {author} {\bibfnamefont {V.~N.}\ \bibnamefont
  {Bogomolov}}\ and\ \bibinfo {author} {\bibfnamefont {D.~N.}\ \bibnamefont
  {Mirlin}},\ }\bibfield  {title} {\bibinfo {title} {Optical absorption by
  polarons in rutile (tio2) single crystals},\ }\href
  {https://doi.org/https://doi.org/10.1002/pssb.19680270144} {\bibfield
  {journal} {\bibinfo  {journal} {physica status solidi (b)}\ }\textbf
  {\bibinfo {volume} {27}},\ \bibinfo {pages} {443} (\bibinfo {year} {1968})},\
  \Eprint
  {https://arxiv.org/abs/https://onlinelibrary.wiley.com/doi/pdf/10.1002/pssb.19680270144}
  {https://onlinelibrary.wiley.com/doi/pdf/10.1002/pssb.19680270144}
  \BibitemShut {NoStop}%
\bibitem [{\citenamefont {Franchini}\ \emph {et~al.}(2021)\citenamefont
  {Franchini}, \citenamefont {Reticcioli}, \citenamefont {Setvin},\ and\
  \citenamefont {Diebold}}]{franchini_polarons_2021}%
  \BibitemOpen
  \bibfield  {author} {\bibinfo {author} {\bibfnamefont {C.}~\bibnamefont
  {Franchini}}, \bibinfo {author} {\bibfnamefont {M.}~\bibnamefont
  {Reticcioli}}, \bibinfo {author} {\bibfnamefont {M.}~\bibnamefont {Setvin}},\
  and\ \bibinfo {author} {\bibfnamefont {U.}~\bibnamefont {Diebold}},\
  }\bibfield  {title} {\bibinfo {title} {Polarons in materials},\ }\href
  {https://doi.org/10.1038/s41578-021-00289-w} {\bibfield  {journal} {\bibinfo
  {journal} {Nature Reviews Materials}\ }\textbf {\bibinfo {volume} {6}},\
  \bibinfo {pages} {560} (\bibinfo {year} {2021})}\BibitemShut {NoStop}%
\bibitem [{\citenamefont {Crevecoeur}\ and\ \citenamefont {{De
  Wit}}(1970)}]{CREVECOEUR1970783}%
  \BibitemOpen
  \bibfield  {author} {\bibinfo {author} {\bibfnamefont {C.}~\bibnamefont
  {Crevecoeur}}\ and\ \bibinfo {author} {\bibfnamefont {H.}~\bibnamefont {{De
  Wit}}},\ }\bibfield  {title} {\bibinfo {title} {Electrical conductivity of li
  doped mno},\ }\href
  {https://doi.org/https://doi.org/10.1016/0022-3697(70)90212-X} {\bibfield
  {journal} {\bibinfo  {journal} {Journal of Physics and Chemistry of Solids}\
  }\textbf {\bibinfo {volume} {31}},\ \bibinfo {pages} {783} (\bibinfo {year}
  {1970})}\BibitemShut {NoStop}%
\bibitem [{\citenamefont {Stoneham}\ \emph {et~al.}(2007)\citenamefont
  {Stoneham}, \citenamefont {Gavartin}, \citenamefont {Shluger}, \citenamefont
  {Kimmel}, \citenamefont {Muñoz~Ramo}, \citenamefont {Rønnow}, \citenamefont
  {Aeppli},\ and\ \citenamefont {Renner}}]{Stoneham_2007}%
  \BibitemOpen
  \bibfield  {author} {\bibinfo {author} {\bibfnamefont {A.~M.}\ \bibnamefont
  {Stoneham}}, \bibinfo {author} {\bibfnamefont {J.}~\bibnamefont {Gavartin}},
  \bibinfo {author} {\bibfnamefont {A.~L.}\ \bibnamefont {Shluger}}, \bibinfo
  {author} {\bibfnamefont {A.~V.}\ \bibnamefont {Kimmel}}, \bibinfo {author}
  {\bibfnamefont {D.}~\bibnamefont {Muñoz~Ramo}}, \bibinfo {author}
  {\bibfnamefont {H.~M.}\ \bibnamefont {Rønnow}}, \bibinfo {author}
  {\bibfnamefont {G.}~\bibnamefont {Aeppli}},\ and\ \bibinfo {author}
  {\bibfnamefont {C.}~\bibnamefont {Renner}},\ }\bibfield  {title} {\bibinfo
  {title} {Trapping, self-trapping and the polaron family},\ }\href
  {https://doi.org/10.1088/0953-8984/19/25/255208} {\bibfield  {journal}
  {\bibinfo  {journal} {Journal of Physics: Condensed Matter}\ }\textbf
  {\bibinfo {volume} {19}},\ \bibinfo {pages} {255208} (\bibinfo {year}
  {2007})}\BibitemShut {NoStop}%
\bibitem [{\citenamefont {Coropceanu}\ \emph {et~al.}(2007)\citenamefont
  {Coropceanu}, \citenamefont {Cornil}, \citenamefont {da~Silva~Filho},
  \citenamefont {Olivier}, \citenamefont {Silbey},\ and\ \citenamefont
  {Brédas}}]{coropceanu_charge_2007}%
  \BibitemOpen
  \bibfield  {author} {\bibinfo {author} {\bibfnamefont {V.}~\bibnamefont
  {Coropceanu}}, \bibinfo {author} {\bibfnamefont {J.}~\bibnamefont {Cornil}},
  \bibinfo {author} {\bibfnamefont {D.~A.}\ \bibnamefont {da~Silva~Filho}},
  \bibinfo {author} {\bibfnamefont {Y.}~\bibnamefont {Olivier}}, \bibinfo
  {author} {\bibfnamefont {R.}~\bibnamefont {Silbey}},\ and\ \bibinfo {author}
  {\bibfnamefont {J.-L.}\ \bibnamefont {Brédas}},\ }\bibfield  {title}
  {\bibinfo {title} {Charge {Transport} in {Organic} {Semiconductors}},\ }\href
  {https://doi.org/10.1021/cr050140x} {\bibfield  {journal} {\bibinfo
  {journal} {Chemical Reviews}\ }\textbf {\bibinfo {volume} {107}},\ \bibinfo
  {pages} {926} (\bibinfo {year} {2007})},\ \bibinfo {note} {publisher:
  American Chemical Society}\BibitemShut {NoStop}%
\bibitem [{\citenamefont {Zhugayevych}\ and\ \citenamefont
  {Tretiak}(2015)}]{WOS:000352259800014}%
  \BibitemOpen
  \bibfield  {author} {\bibinfo {author} {\bibfnamefont {A.}~\bibnamefont
  {Zhugayevych}}\ and\ \bibinfo {author} {\bibfnamefont {S.}~\bibnamefont
  {Tretiak}},\ }\bibfield  {title} {\bibinfo {title} {Theoretical description
  of structural and electronic properties of organic photovoltaic materials},\
  }in\ \href {https://doi.org/10.1146/annurev-physchem-040214-121440} {\emph
  {\bibinfo {booktitle} {ANNUAL REVIEW OF PHYSICAL CHEMISTRY, VOL 66}}},\
  \bibinfo {series} {Annual Review of Physical Chemistry}, Vol.~\bibinfo
  {volume} {66},\ \bibinfo {editor} {edited by\ \bibinfo {editor}
  {\bibfnamefont {M.}~\bibnamefont {Johnson}}\ and\ \bibinfo {editor}
  {\bibfnamefont {T.}~\bibnamefont {Martinez}}}\ (\bibinfo {year} {2015})\ pp.\
  \bibinfo {pages} {305+}\BibitemShut {NoStop}%
\bibitem [{\citenamefont {De~Sio}\ \emph {et~al.}(2016)\citenamefont {De~Sio},
  \citenamefont {Troiani}, \citenamefont {Maiuri}, \citenamefont {Réhault},
  \citenamefont {Sommer}, \citenamefont {Lim}, \citenamefont {Huelga},
  \citenamefont {Plenio}, \citenamefont {Rozzi}, \citenamefont {Cerullo},
  \citenamefont {Molinari},\ and\ \citenamefont
  {Lienau}}]{de_sio_tracking_2016}%
  \BibitemOpen
  \bibfield  {author} {\bibinfo {author} {\bibfnamefont {A.}~\bibnamefont
  {De~Sio}}, \bibinfo {author} {\bibfnamefont {F.}~\bibnamefont {Troiani}},
  \bibinfo {author} {\bibfnamefont {M.}~\bibnamefont {Maiuri}}, \bibinfo
  {author} {\bibfnamefont {J.}~\bibnamefont {Réhault}}, \bibinfo {author}
  {\bibfnamefont {E.}~\bibnamefont {Sommer}}, \bibinfo {author} {\bibfnamefont
  {J.}~\bibnamefont {Lim}}, \bibinfo {author} {\bibfnamefont {S.~F.}\
  \bibnamefont {Huelga}}, \bibinfo {author} {\bibfnamefont {M.~B.}\
  \bibnamefont {Plenio}}, \bibinfo {author} {\bibfnamefont {C.~A.}\
  \bibnamefont {Rozzi}}, \bibinfo {author} {\bibfnamefont {G.}~\bibnamefont
  {Cerullo}}, \bibinfo {author} {\bibfnamefont {E.}~\bibnamefont {Molinari}},\
  and\ \bibinfo {author} {\bibfnamefont {C.}~\bibnamefont {Lienau}},\
  }\bibfield  {title} {\bibinfo {title} {Tracking the coherent generation of
  polaron pairs in conjugated polymers},\ }\href
  {https://doi.org/10.1038/ncomms13742} {\bibfield  {journal} {\bibinfo
  {journal} {Nature Communications}\ }\textbf {\bibinfo {volume} {7}},\
  \bibinfo {pages} {13742} (\bibinfo {year} {2016})}\BibitemShut {NoStop}%
\bibitem [{\citenamefont {Kaminski}\ and\ \citenamefont
  {Das~Sarma}(2002)}]{PhysRevLett.88.247202}%
  \BibitemOpen
  \bibfield  {author} {\bibinfo {author} {\bibfnamefont {A.}~\bibnamefont
  {Kaminski}}\ and\ \bibinfo {author} {\bibfnamefont {S.}~\bibnamefont
  {Das~Sarma}},\ }\bibfield  {title} {\bibinfo {title} {Polaron percolation in
  diluted magnetic semiconductors},\ }\href
  {https://doi.org/10.1103/PhysRevLett.88.247202} {\bibfield  {journal}
  {\bibinfo  {journal} {Phys. Rev. Lett.}\ }\textbf {\bibinfo {volume} {88}},\
  \bibinfo {pages} {247202} (\bibinfo {year} {2002})}\BibitemShut {NoStop}%
\bibitem [{\citenamefont {Teresa}\ \emph {et~al.}(1997)\citenamefont {Teresa},
  \citenamefont {Ibarra}, \citenamefont {Algarabel}, \citenamefont {Ritter},
  \citenamefont {Marquina}, \citenamefont {Blasco}, \citenamefont {García},
  \citenamefont {del Moral},\ and\ \citenamefont
  {Arnold}}]{teresa_evidence_1997}%
  \BibitemOpen
  \bibfield  {author} {\bibinfo {author} {\bibfnamefont {J.~M.~D.}\
  \bibnamefont {Teresa}}, \bibinfo {author} {\bibfnamefont {M.~R.}\
  \bibnamefont {Ibarra}}, \bibinfo {author} {\bibfnamefont {P.~A.}\
  \bibnamefont {Algarabel}}, \bibinfo {author} {\bibfnamefont {C.}~\bibnamefont
  {Ritter}}, \bibinfo {author} {\bibfnamefont {C.}~\bibnamefont {Marquina}},
  \bibinfo {author} {\bibfnamefont {J.}~\bibnamefont {Blasco}}, \bibinfo
  {author} {\bibfnamefont {J.}~\bibnamefont {García}}, \bibinfo {author}
  {\bibfnamefont {A.}~\bibnamefont {del Moral}},\ and\ \bibinfo {author}
  {\bibfnamefont {Z.}~\bibnamefont {Arnold}},\ }\bibfield  {title} {\bibinfo
  {title} {Evidence for magnetic polarons in the magnetoresistive
  perovskites},\ }\href {https://doi.org/10.1038/386256a0} {\bibfield
  {journal} {\bibinfo  {journal} {Nature}\ }\textbf {\bibinfo {volume} {386}},\
  \bibinfo {pages} {256} (\bibinfo {year} {1997})}\BibitemShut {NoStop}%
\bibitem [{\citenamefont {Daoud-Aladine}\ \emph {et~al.}(2002)\citenamefont
  {Daoud-Aladine}, \citenamefont {Rodr\'{\i}guez-Carvajal}, \citenamefont
  {Pinsard-Gaudart}, \citenamefont {Fern\'andez-D\'{\i}az},\ and\ \citenamefont
  {Revcolevschi}}]{PhysRevLett.89.097205}%
  \BibitemOpen
  \bibfield  {author} {\bibinfo {author} {\bibfnamefont {A.}~\bibnamefont
  {Daoud-Aladine}}, \bibinfo {author} {\bibfnamefont {J.}~\bibnamefont
  {Rodr\'{\i}guez-Carvajal}}, \bibinfo {author} {\bibfnamefont
  {L.}~\bibnamefont {Pinsard-Gaudart}}, \bibinfo {author} {\bibfnamefont
  {M.~T.}\ \bibnamefont {Fern\'andez-D\'{\i}az}},\ and\ \bibinfo {author}
  {\bibfnamefont {A.}~\bibnamefont {Revcolevschi}},\ }\bibfield  {title}
  {\bibinfo {title} {Zener polaron ordering in half-doped manganites},\ }\href
  {https://doi.org/10.1103/PhysRevLett.89.097205} {\bibfield  {journal}
  {\bibinfo  {journal} {Phys. Rev. Lett.}\ }\textbf {\bibinfo {volume} {89}},\
  \bibinfo {pages} {097205} (\bibinfo {year} {2002})}\BibitemShut {NoStop}%
\bibitem [{\citenamefont {Yamada}\ \emph {et~al.}(1996)\citenamefont {Yamada},
  \citenamefont {Hino}, \citenamefont {Nohdo}, \citenamefont {Kanao},
  \citenamefont {Inami},\ and\ \citenamefont {Katano}}]{PhysRevLett.77.904}%
  \BibitemOpen
  \bibfield  {author} {\bibinfo {author} {\bibfnamefont {Y.}~\bibnamefont
  {Yamada}}, \bibinfo {author} {\bibfnamefont {O.}~\bibnamefont {Hino}},
  \bibinfo {author} {\bibfnamefont {S.}~\bibnamefont {Nohdo}}, \bibinfo
  {author} {\bibfnamefont {R.}~\bibnamefont {Kanao}}, \bibinfo {author}
  {\bibfnamefont {T.}~\bibnamefont {Inami}},\ and\ \bibinfo {author}
  {\bibfnamefont {S.}~\bibnamefont {Katano}},\ }\bibfield  {title} {\bibinfo
  {title} {Polaron ordering in low-doping
  ${\mathrm{la}}_{1\ensuremath{-}\mathit{x}}{\mathrm{sr}}_{\mathit{x}}{\mathrm{mno}}_{3}$},\
  }\href {https://doi.org/10.1103/PhysRevLett.77.904} {\bibfield  {journal}
  {\bibinfo  {journal} {Phys. Rev. Lett.}\ }\textbf {\bibinfo {volume} {77}},\
  \bibinfo {pages} {904} (\bibinfo {year} {1996})}\BibitemShut {NoStop}%
\bibitem [{\citenamefont {Zhao}\ \emph {et~al.}(1997)\citenamefont {Zhao},
  \citenamefont {Hunt}, \citenamefont {Keller},\ and\ \citenamefont
  {Müller}}]{zhao_evidence_1997}%
  \BibitemOpen
  \bibfield  {author} {\bibinfo {author} {\bibfnamefont {G.-m.}\ \bibnamefont
  {Zhao}}, \bibinfo {author} {\bibfnamefont {M.~B.}\ \bibnamefont {Hunt}},
  \bibinfo {author} {\bibfnamefont {H.}~\bibnamefont {Keller}},\ and\ \bibinfo
  {author} {\bibfnamefont {K.~A.}\ \bibnamefont {Müller}},\ }\bibfield
  {title} {\bibinfo {title} {Evidence for polaronic supercarriers in the copper
  oxide superconductors {La2}–{xSrxCuO4}},\ }\href
  {https://doi.org/10.1038/385236a0} {\bibfield  {journal} {\bibinfo  {journal}
  {Nature}\ }\textbf {\bibinfo {volume} {385}},\ \bibinfo {pages} {236}
  (\bibinfo {year} {1997})}\BibitemShut {NoStop}%
\bibitem [{\citenamefont {Cortecchia}\ \emph {et~al.}(2017)\citenamefont
  {Cortecchia}, \citenamefont {Yin}, \citenamefont {Bruno}, \citenamefont {Lo},
  \citenamefont {Gurzadyan}, \citenamefont {Mhaisalkar}, \citenamefont
  {Brédas},\ and\ \citenamefont {Soci}}]{C7TC00366H}%
  \BibitemOpen
  \bibfield  {author} {\bibinfo {author} {\bibfnamefont {D.}~\bibnamefont
  {Cortecchia}}, \bibinfo {author} {\bibfnamefont {J.}~\bibnamefont {Yin}},
  \bibinfo {author} {\bibfnamefont {A.}~\bibnamefont {Bruno}}, \bibinfo
  {author} {\bibfnamefont {S.-Z.~A.}\ \bibnamefont {Lo}}, \bibinfo {author}
  {\bibfnamefont {G.~G.}\ \bibnamefont {Gurzadyan}}, \bibinfo {author}
  {\bibfnamefont {S.}~\bibnamefont {Mhaisalkar}}, \bibinfo {author}
  {\bibfnamefont {J.-L.}\ \bibnamefont {Brédas}},\ and\ \bibinfo {author}
  {\bibfnamefont {C.}~\bibnamefont {Soci}},\ }\bibfield  {title} {\bibinfo
  {title} {Polaron self-localization in white-light emitting hybrid
  perovskites},\ }\href {https://doi.org/10.1039/C7TC00366H} {\bibfield
  {journal} {\bibinfo  {journal} {J. Mater. Chem. C}\ }\textbf {\bibinfo
  {volume} {5}},\ \bibinfo {pages} {2771} (\bibinfo {year} {2017})}\BibitemShut
  {NoStop}%
\bibitem [{\citenamefont {Miyata}\ \emph {et~al.}(2017)\citenamefont {Miyata},
  \citenamefont {Meggiolaro}, \citenamefont {Trinh}, \citenamefont {Joshi},
  \citenamefont {Mosconi}, \citenamefont {Jones}, \citenamefont {Angelis},\
  and\ \citenamefont {Zhu}}]{doi:10.1126/sciadv.1701217}%
  \BibitemOpen
  \bibfield  {author} {\bibinfo {author} {\bibfnamefont {K.}~\bibnamefont
  {Miyata}}, \bibinfo {author} {\bibfnamefont {D.}~\bibnamefont {Meggiolaro}},
  \bibinfo {author} {\bibfnamefont {M.~T.}\ \bibnamefont {Trinh}}, \bibinfo
  {author} {\bibfnamefont {P.~P.}\ \bibnamefont {Joshi}}, \bibinfo {author}
  {\bibfnamefont {E.}~\bibnamefont {Mosconi}}, \bibinfo {author} {\bibfnamefont
  {S.~C.}\ \bibnamefont {Jones}}, \bibinfo {author} {\bibfnamefont {F.~D.}\
  \bibnamefont {Angelis}},\ and\ \bibinfo {author} {\bibfnamefont {X.-Y.}\
  \bibnamefont {Zhu}},\ }\bibfield  {title} {\bibinfo {title} {Large polarons
  in lead halide perovskites},\ }\href {https://doi.org/10.1126/sciadv.1701217}
  {\bibfield  {journal} {\bibinfo  {journal} {Science Advances}\ }\textbf
  {\bibinfo {volume} {3}},\ \bibinfo {pages} {e1701217} (\bibinfo {year}
  {2017})},\ \Eprint
  {https://arxiv.org/abs/https://www.science.org/doi/pdf/10.1126/sciadv.1701217}
  {https://www.science.org/doi/pdf/10.1126/sciadv.1701217} \BibitemShut
  {NoStop}%
\bibitem [{\citenamefont {Chen}\ \emph {et~al.}(2018)\citenamefont {Chen},
  \citenamefont {Wang},\ and\ \citenamefont {Peeters}}]{10.1063/1.5025907}%
  \BibitemOpen
  \bibfield  {author} {\bibinfo {author} {\bibfnamefont {Q.}~\bibnamefont
  {Chen}}, \bibinfo {author} {\bibfnamefont {W.}~\bibnamefont {Wang}},\ and\
  \bibinfo {author} {\bibfnamefont {F.~M.}\ \bibnamefont {Peeters}},\
  }\bibfield  {title} {\bibinfo {title} {Magneto-polarons in monolayer
  transition-metal dichalcogenides},\ }\href
  {https://doi.org/10.1063/1.5025907} {\bibfield  {journal} {\bibinfo
  {journal} {Journal of Applied Physics}\ }\textbf {\bibinfo {volume} {123}},\
  \bibinfo {pages} {214303} (\bibinfo {year} {2018})},\ \Eprint
  {https://arxiv.org/abs/https://pubs.aip.org/aip/jap/article-pdf/doi/10.1063/1.5025907/15212668/214303\_1\_online.pdf}
  {https://pubs.aip.org/aip/jap/article-pdf/doi/10.1063/1.5025907/15212668/214303\_1\_online.pdf}
  \BibitemShut {NoStop}%
\bibitem [{\citenamefont {Kang}\ \emph {et~al.}(2018)\citenamefont {Kang},
  \citenamefont {Jung}, \citenamefont {Shin}, \citenamefont {Sohn},
  \citenamefont {Ryu}, \citenamefont {Kim}, \citenamefont {Hoesch},\ and\
  \citenamefont {Kim}}]{WOS:000439573400010}%
  \BibitemOpen
  \bibfield  {author} {\bibinfo {author} {\bibfnamefont {M.}~\bibnamefont
  {Kang}}, \bibinfo {author} {\bibfnamefont {S.~W.}\ \bibnamefont {Jung}},
  \bibinfo {author} {\bibfnamefont {W.~J.}\ \bibnamefont {Shin}}, \bibinfo
  {author} {\bibfnamefont {Y.}~\bibnamefont {Sohn}}, \bibinfo {author}
  {\bibfnamefont {S.~H.}\ \bibnamefont {Ryu}}, \bibinfo {author} {\bibfnamefont
  {T.~K.}\ \bibnamefont {Kim}}, \bibinfo {author} {\bibfnamefont
  {M.}~\bibnamefont {Hoesch}},\ and\ \bibinfo {author} {\bibfnamefont {K.~S.}\
  \bibnamefont {Kim}},\ }\bibfield  {title} {\bibinfo {title} {Holstein polaron
  in a valley-degenerate two-dimensional semiconductor},\ }\href
  {https://doi.org/10.1038/s41563-018-0092-7} {\bibfield  {journal} {\bibinfo
  {journal} {NATURE MATERIALS}\ }\textbf {\bibinfo {volume} {17}},\ \bibinfo
  {pages} {676+} (\bibinfo {year} {2018})}\BibitemShut {NoStop}%
\bibitem [{\citenamefont {McKenna}\ \emph {et~al.}(2012)\citenamefont
  {McKenna}, \citenamefont {Wolf}, \citenamefont {Shluger}, \citenamefont
  {Lany},\ and\ \citenamefont {Zunger}}]{PhysRevLett.108.116403}%
  \BibitemOpen
  \bibfield  {author} {\bibinfo {author} {\bibfnamefont {K.~P.}\ \bibnamefont
  {McKenna}}, \bibinfo {author} {\bibfnamefont {M.~J.}\ \bibnamefont {Wolf}},
  \bibinfo {author} {\bibfnamefont {A.~L.}\ \bibnamefont {Shluger}}, \bibinfo
  {author} {\bibfnamefont {S.}~\bibnamefont {Lany}},\ and\ \bibinfo {author}
  {\bibfnamefont {A.}~\bibnamefont {Zunger}},\ }\bibfield  {title} {\bibinfo
  {title} {Two-dimensional polaronic behavior in the binary oxides
  $m\mathrm{\text{\ensuremath{-}}}{\mathrm{hfo}}_{2}$ and
  $m\mathrm{\text{\ensuremath{-}}}{\mathrm{zro}}_{2}$},\ }\href
  {https://doi.org/10.1103/PhysRevLett.108.116403} {\bibfield  {journal}
  {\bibinfo  {journal} {Phys. Rev. Lett.}\ }\textbf {\bibinfo {volume} {108}},\
  \bibinfo {pages} {116403} (\bibinfo {year} {2012})}\BibitemShut {NoStop}%
\bibitem [{\citenamefont {Krotz}\ and\ \citenamefont
  {Tempelaar}(2022)}]{10.1063/5.0076070}%
  \BibitemOpen
  \bibfield  {author} {\bibinfo {author} {\bibfnamefont {A.}~\bibnamefont
  {Krotz}}\ and\ \bibinfo {author} {\bibfnamefont {R.}~\bibnamefont
  {Tempelaar}},\ }\bibfield  {title} {\bibinfo {title} {A reciprocal-space
  formulation of surface hopping},\ }\href {https://doi.org/10.1063/5.0076070}
  {\bibfield  {journal} {\bibinfo  {journal} {The Journal of Chemical Physics}\
  }\textbf {\bibinfo {volume} {156}},\ \bibinfo {pages} {024105} (\bibinfo
  {year} {2022})},\ \Eprint
  {https://arxiv.org/abs/https://pubs.aip.org/aip/jcp/article-pdf/doi/10.1063/5.0076070/16532057/024105\_1\_online.pdf}
  {https://pubs.aip.org/aip/jcp/article-pdf/doi/10.1063/5.0076070/16532057/024105\_1\_online.pdf}
  \BibitemShut {NoStop}%
\bibitem [{\citenamefont {Morita}\ \emph {et~al.}(2023)\citenamefont {Morita},
  \citenamefont {Golomb}, \citenamefont {Rivera},\ and\ \citenamefont
  {Walsh}}]{morita_models_2023}%
  \BibitemOpen
  \bibfield  {author} {\bibinfo {author} {\bibfnamefont {K.}~\bibnamefont
  {Morita}}, \bibinfo {author} {\bibfnamefont {M.~J.}\ \bibnamefont {Golomb}},
  \bibinfo {author} {\bibfnamefont {M.}~\bibnamefont {Rivera}},\ and\ \bibinfo
  {author} {\bibfnamefont {A.}~\bibnamefont {Walsh}},\ }\bibfield  {title}
  {\bibinfo {title} {Models of {Polaron} {Transport} in {Inorganic} and
  {Hybrid} {Organic}–{Inorganic} {Titanium} {Oxides}},\ }\href
  {https://doi.org/10.1021/acs.chemmater.3c00322} {\bibfield  {journal}
  {\bibinfo  {journal} {Chemistry of Materials}\ }\textbf {\bibinfo {volume}
  {35}},\ \bibinfo {pages} {3652} (\bibinfo {year} {2023})},\ \bibinfo {note}
  {publisher: American Chemical Society}\BibitemShut {NoStop}%
\bibitem [{\citenamefont {Holstein}(1959)}]{HOLSTEIN}%
  \BibitemOpen
  \bibfield  {author} {\bibinfo {author} {\bibfnamefont {T.}~\bibnamefont
  {Holstein}},\ }\bibfield  {title} {\bibinfo {title} {Studies of polaron
  motion: Part i. the molecular-crystal model},\ }\href
  {https://doi.org/https://doi.org/10.1016/0003-4916(59)90002-8} {\bibfield
  {journal} {\bibinfo  {journal} {Annals of Physics}\ }\textbf {\bibinfo
  {volume} {8}},\ \bibinfo {pages} {325} (\bibinfo {year} {1959})}\BibitemShut
  {NoStop}%
\bibitem [{\citenamefont {Meyer}\ \emph {et~al.}(2002)\citenamefont {Meyer},
  \citenamefont {Hewson},\ and\ \citenamefont {Bulla}}]{PhysRevLett.89.196401}%
  \BibitemOpen
  \bibfield  {author} {\bibinfo {author} {\bibfnamefont {D.}~\bibnamefont
  {Meyer}}, \bibinfo {author} {\bibfnamefont {A.~C.}\ \bibnamefont {Hewson}},\
  and\ \bibinfo {author} {\bibfnamefont {R.}~\bibnamefont {Bulla}},\ }\bibfield
   {title} {\bibinfo {title} {Gap formation and soft phonon mode in the
  holstein model},\ }\href {https://doi.org/10.1103/PhysRevLett.89.196401}
  {\bibfield  {journal} {\bibinfo  {journal} {Phys. Rev. Lett.}\ }\textbf
  {\bibinfo {volume} {89}},\ \bibinfo {pages} {196401} (\bibinfo {year}
  {2002})}\BibitemShut {NoStop}%
\bibitem [{\citenamefont {Wellein}\ and\ \citenamefont
  {Fehske}(1997)}]{PhysRevB.56.4513}%
  \BibitemOpen
  \bibfield  {author} {\bibinfo {author} {\bibfnamefont {G.}~\bibnamefont
  {Wellein}}\ and\ \bibinfo {author} {\bibfnamefont {H.}~\bibnamefont
  {Fehske}},\ }\bibfield  {title} {\bibinfo {title} {Polaron band formation in
  the holstein model},\ }\href {https://doi.org/10.1103/PhysRevB.56.4513}
  {\bibfield  {journal} {\bibinfo  {journal} {Phys. Rev. B}\ }\textbf {\bibinfo
  {volume} {56}},\ \bibinfo {pages} {4513} (\bibinfo {year}
  {1997})}\BibitemShut {NoStop}%
\bibitem [{\citenamefont {Freericks}\ \emph {et~al.}(1993)\citenamefont
  {Freericks}, \citenamefont {Jarrell},\ and\ \citenamefont
  {Scalapino}}]{PhysRevB.48.6302}%
  \BibitemOpen
  \bibfield  {author} {\bibinfo {author} {\bibfnamefont {J.~K.}\ \bibnamefont
  {Freericks}}, \bibinfo {author} {\bibfnamefont {M.}~\bibnamefont {Jarrell}},\
  and\ \bibinfo {author} {\bibfnamefont {D.~J.}\ \bibnamefont {Scalapino}},\
  }\bibfield  {title} {\bibinfo {title} {Holstein model in infinite
  dimensions},\ }\href {https://doi.org/10.1103/PhysRevB.48.6302} {\bibfield
  {journal} {\bibinfo  {journal} {Phys. Rev. B}\ }\textbf {\bibinfo {volume}
  {48}},\ \bibinfo {pages} {6302} (\bibinfo {year} {1993})}\BibitemShut
  {NoStop}%
\bibitem [{\citenamefont {Zhang}\ \emph {et~al.}(1999)\citenamefont {Zhang},
  \citenamefont {Jeckelmann},\ and\ \citenamefont {White}}]{PhysRevB.60.14092}%
  \BibitemOpen
  \bibfield  {author} {\bibinfo {author} {\bibfnamefont {C.}~\bibnamefont
  {Zhang}}, \bibinfo {author} {\bibfnamefont {E.}~\bibnamefont {Jeckelmann}},\
  and\ \bibinfo {author} {\bibfnamefont {S.~R.}\ \bibnamefont {White}},\
  }\bibfield  {title} {\bibinfo {title} {Dynamical properties of the
  one-dimensional holstein model},\ }\href
  {https://doi.org/10.1103/PhysRevB.60.14092} {\bibfield  {journal} {\bibinfo
  {journal} {Phys. Rev. B}\ }\textbf {\bibinfo {volume} {60}},\ \bibinfo
  {pages} {14092} (\bibinfo {year} {1999})}\BibitemShut {NoStop}%
\bibitem [{\citenamefont {ten Brink}\ \emph {et~al.}(2022)\citenamefont {ten
  Brink}, \citenamefont {Gr\"{a}ber}, \citenamefont {Hopjan}, \citenamefont
  {Jansen}, \citenamefont {Stolpp}, \citenamefont {Heidrich-Meisner},\ and\
  \citenamefont {Bl\"{o}chl}}]{Holstein_Brink}%
  \BibitemOpen
  \bibfield  {author} {\bibinfo {author} {\bibfnamefont {M.}~\bibnamefont {ten
  Brink}}, \bibinfo {author} {\bibfnamefont {S.}~\bibnamefont {Gr\"{a}ber}},
  \bibinfo {author} {\bibfnamefont {M.}~\bibnamefont {Hopjan}}, \bibinfo
  {author} {\bibfnamefont {D.}~\bibnamefont {Jansen}}, \bibinfo {author}
  {\bibfnamefont {J.}~\bibnamefont {Stolpp}}, \bibinfo {author} {\bibfnamefont
  {F.}~\bibnamefont {Heidrich-Meisner}},\ and\ \bibinfo {author} {\bibfnamefont
  {P.~E.}\ \bibnamefont {Bl\"{o}chl}},\ }\bibfield  {title} {\bibinfo {title}
  {Real-time non-adiabatic dynamics in the one-dimensional holstein model:
  Trajectory-based vs exact methods},\ }\bibfield  {journal} {\bibinfo
  {journal} {The Journal of Chemical Physics}\ }\textbf {\bibinfo {volume}
  {156}},\ \href {https://doi.org/10.1063/5.0092063} {10.1063/5.0092063}
  (\bibinfo {year} {2022})\BibitemShut {NoStop}%
\bibitem [{\citenamefont {Murakami}\ \emph {et~al.}(2013)\citenamefont
  {Murakami}, \citenamefont {Werner}, \citenamefont {Tsuji},\ and\
  \citenamefont {Aoki}}]{PhysRevB.88.125126}%
  \BibitemOpen
  \bibfield  {author} {\bibinfo {author} {\bibfnamefont {Y.}~\bibnamefont
  {Murakami}}, \bibinfo {author} {\bibfnamefont {P.}~\bibnamefont {Werner}},
  \bibinfo {author} {\bibfnamefont {N.}~\bibnamefont {Tsuji}},\ and\ \bibinfo
  {author} {\bibfnamefont {H.}~\bibnamefont {Aoki}},\ }\bibfield  {title}
  {\bibinfo {title} {Ordered phases in the holstein-hubbard model: Interplay of
  strong coulomb interaction and electron-phonon coupling},\ }\href
  {https://doi.org/10.1103/PhysRevB.88.125126} {\bibfield  {journal} {\bibinfo
  {journal} {Phys. Rev. B}\ }\textbf {\bibinfo {volume} {88}},\ \bibinfo
  {pages} {125126} (\bibinfo {year} {2013})}\BibitemShut {NoStop}%
\bibitem [{\citenamefont {Backes}\ \emph {et~al.}(2023)\citenamefont {Backes},
  \citenamefont {Murakami}, \citenamefont {Sakai},\ and\ \citenamefont
  {Arita}}]{PhysRevB.107.165155}%
  \BibitemOpen
  \bibfield  {author} {\bibinfo {author} {\bibfnamefont {S.}~\bibnamefont
  {Backes}}, \bibinfo {author} {\bibfnamefont {Y.}~\bibnamefont {Murakami}},
  \bibinfo {author} {\bibfnamefont {S.}~\bibnamefont {Sakai}},\ and\ \bibinfo
  {author} {\bibfnamefont {R.}~\bibnamefont {Arita}},\ }\bibfield  {title}
  {\bibinfo {title} {Dynamical mean-field theory for the hubbard-holstein model
  on a quantum device},\ }\href {https://doi.org/10.1103/PhysRevB.107.165155}
  {\bibfield  {journal} {\bibinfo  {journal} {Phys. Rev. B}\ }\textbf {\bibinfo
  {volume} {107}},\ \bibinfo {pages} {165155} (\bibinfo {year}
  {2023})}\BibitemShut {NoStop}%
\bibitem [{\citenamefont {Kemper}\ \emph {et~al.}(2013)\citenamefont {Kemper},
  \citenamefont {Moritz}, \citenamefont {Freericks},\ and\ \citenamefont
  {Devereaux}}]{WOS:000314516500003}%
  \BibitemOpen
  \bibfield  {author} {\bibinfo {author} {\bibfnamefont {A.~F.}\ \bibnamefont
  {Kemper}}, \bibinfo {author} {\bibfnamefont {B.}~\bibnamefont {Moritz}},
  \bibinfo {author} {\bibfnamefont {J.~K.}\ \bibnamefont {Freericks}},\ and\
  \bibinfo {author} {\bibfnamefont {T.~P.}\ \bibnamefont {Devereaux}},\
  }\bibfield  {title} {\bibinfo {title} {Theoretical description of high-order
  harmonic generation in solids},\ }\bibfield  {journal} {\bibinfo  {journal}
  {NEW JOURNAL OF PHYSICS}\ }\textbf {\bibinfo {volume} {15}},\ \href
  {https://doi.org/10.1088/1367-2630/15/2/023003}
  {10.1088/1367-2630/15/2/023003} (\bibinfo {year} {2013})\BibitemShut
  {NoStop}%
\bibitem [{\citenamefont {Tsuji}\ \emph {et~al.}(2016)\citenamefont {Tsuji},
  \citenamefont {Murakami},\ and\ \citenamefont {Aoki}}]{WOS:000391009300002}%
  \BibitemOpen
  \bibfield  {author} {\bibinfo {author} {\bibfnamefont {N.}~\bibnamefont
  {Tsuji}}, \bibinfo {author} {\bibfnamefont {Y.}~\bibnamefont {Murakami}},\
  and\ \bibinfo {author} {\bibfnamefont {H.}~\bibnamefont {Aoki}},\ }\bibfield
  {title} {\bibinfo {title} {Nonlinear light-higgs coupling in superconductors
  beyond bcs: Effects of the retarded phonon-mediated interaction},\ }\bibfield
   {journal} {\bibinfo  {journal} {PHYSICAL REVIEW B}\ }\textbf {\bibinfo
  {volume} {94}},\ \href {https://doi.org/10.1103/PhysRevB.94.224519}
  {10.1103/PhysRevB.94.224519} (\bibinfo {year} {2016})\BibitemShut {NoStop}%
\bibitem [{\citenamefont {Sota}\ \emph {et~al.}(2015)\citenamefont {Sota},
  \citenamefont {Yunoki},\ and\ \citenamefont
  {Tohyama}}]{doi:10.7566/JPSJ.84.054403}%
  \BibitemOpen
  \bibfield  {author} {\bibinfo {author} {\bibfnamefont {S.}~\bibnamefont
  {Sota}}, \bibinfo {author} {\bibfnamefont {S.}~\bibnamefont {Yunoki}},\ and\
  \bibinfo {author} {\bibfnamefont {T.}~\bibnamefont {Tohyama}},\ }\bibfield
  {title} {\bibinfo {title} {Density-matrix renormalization group study of
  third harmonic generation in one-dimensional mott insulator coupled with
  phonon},\ }\href {https://doi.org/10.7566/JPSJ.84.054403} {\bibfield
  {journal} {\bibinfo  {journal} {Journal of the Physical Society of Japan}\
  }\textbf {\bibinfo {volume} {84}},\ \bibinfo {pages} {054403} (\bibinfo
  {year} {2015})},\ \Eprint
  {https://arxiv.org/abs/https://doi.org/10.7566/JPSJ.84.054403}
  {https://doi.org/10.7566/JPSJ.84.054403} \BibitemShut {NoStop}%
\bibitem [{\citenamefont {Su}\ \emph {et~al.}(1979)\citenamefont {Su},
  \citenamefont {Schrieffer},\ and\ \citenamefont
  {Heeger}}]{PhysRevLett.42.1698}%
  \BibitemOpen
  \bibfield  {author} {\bibinfo {author} {\bibfnamefont {W.~P.}\ \bibnamefont
  {Su}}, \bibinfo {author} {\bibfnamefont {J.~R.}\ \bibnamefont {Schrieffer}},\
  and\ \bibinfo {author} {\bibfnamefont {A.~J.}\ \bibnamefont {Heeger}},\
  }\bibfield  {title} {\bibinfo {title} {Solitons in polyacetylene},\ }\href
  {https://doi.org/10.1103/PhysRevLett.42.1698} {\bibfield  {journal} {\bibinfo
   {journal} {Phys. Rev. Lett.}\ }\textbf {\bibinfo {volume} {42}},\ \bibinfo
  {pages} {1698} (\bibinfo {year} {1979})}\BibitemShut {NoStop}%
\end{thebibliography}%

\appendix

\section{Supplementary material}

\subsection{Relevant Eigenstates}

In the simulation, the most relevant state is the ground state, as the system starts from this state. Additionally, the states with strong transition from the ground state are important. We identify these states by selecting those with the highest values of $\log_{10}(T_\mathrm{gs}^2)$, where the log is to sort the result by order of magnitude. 

\begin{figure}[h!]
    \centering
    \includegraphics[width=1.0\linewidth]{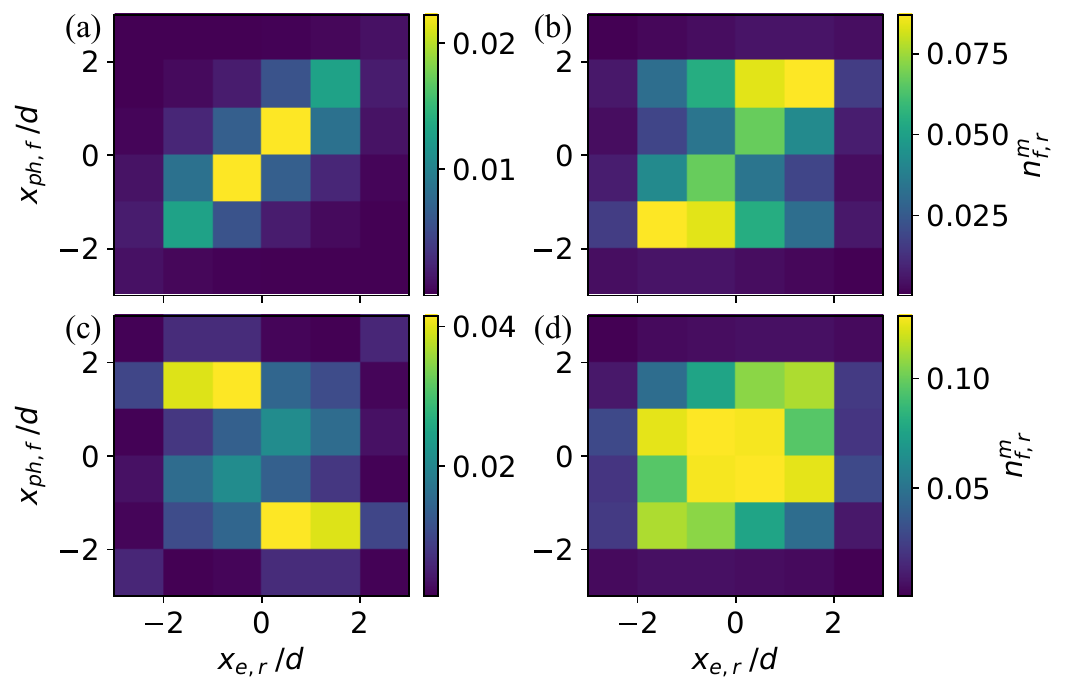}
    \caption{Electron distribution relative to the phonon distribution for the four most relevant states. $f$ and $r$ indicate the $f$-th and $r$-th site, respectively. (a) Ground state ($m=0$). (b) First excited state ($m=1$). (c) $m=7$. (d) $m=12$. Panels (b), (c) and (d) correspond to the states with the highest values of $\log_{10}(T_{gs}^2)$. The same coupling and phonon parameters were used as in Fig. \ref{fig:energy-levels}. }
    \label{fig:relevant_states}
\end{figure}

The inclusion of phonons leads to more complex eigenstates, each characterized by a distinct electron and phonon distribution. Figure \ref{fig:relevant_states}.(a) shows the ground state, where we plot both the electron and phonon distributions. To this end, we calculate the expectation value of the product of the phonon number operator and the electron number operator at different sites:
\begin{align}
    n_{f,r}^m= \bra{\phi_m} \hat n_{ph,f}\hspace{0.1cm} \hat n_{e,r}  \ket{\phi_m}\ ,
\end{align}
where $f$ and $r$  indicate the $f$th and $r$th site, respectively, and $m$ refers to the $m$th eigenstate. 
This quantity illustrates how the electron distribution correlates with the phonon distribution in the eigenstate $\phi_m$. For the ground state, we observe that electrons and phonons are more likely to be together near the center of the lattice. 
The three eigenstates with the higher values of $T_\mathrm{gs}^2$ are as follows: the first excited state ($m=1$) in Fig. \ref{fig:relevant_states}.(b), seventh ($m=7$) in Fig. \ref{fig:relevant_states}.(c), and twelfth ($m=12$) in Fig. \ref{fig:relevant_states}.(d). These states are highlighted in blue in Fig.\ref{fig:energy-levels}.

\end{document}